\documentclass[sort&compress]{elsarticle}
\pdfoutput=1
\usepackage{fullpage}
\usepackage{color}
\usepackage{amsmath}
\usepackage{amssymb}
\usepackage{amsfonts}
\usepackage{graphicx}
\usepackage{hyperref}
\usepackage[normalem]{ulem}

\newcommand{\Navg}{\left\langle N_{\xi} \right\rangle}
\newcommand{\MCSRD}{\rm{SRD}^{mc}}
\newcommand{\MCSRDp}{\rm{SRD}^{mc}_{+}}
\newcommand{\MCSRDm}{\rm{SRD}^{mc}_{-}}
\newcommand{\MFP}{\lambda_{\rm MFP}}

\newcommand{\comments}[1]{}%
\newcommand{\newtext}[1]{\textcolor{blue}{#1}}
\newcommand{\oldtext}[1]{\textcolor{red}{\sout{#1}}}

\renewcommand{\newtext}[1]{\textcolor{black}{#1}}
\renewcommand{\oldtext}[1]{\comments{#1}}

\DeclareMathOperator{\sgn}{sgn}
\DeclareMathOperator{\Tr}{tr}

\begin{document}

\title{Stochastic Rotation Dynamics simulations of wetting multi--phase flows}

\author[dcf]{Thomas Hiller\corref{cor1}\fnref{fn1}}
\ead{THiller@eonerc.rwth-aachen.de}
\author[dcf]{Marta Sanchez de La Lama\fnref{fn2}}
\author[dcf,sub]{Martin Brinkmann}
\cortext[cor1]{Corresponding author}
\address[dcf]{Dynamics of Complex Fluids, Max Planck Institute for Dynamics and Self--Organization, Am Fassberg 17, 37077 G\"{o}ttingen, Germany}
\address[sub]{Department of Experimental Physics, Universit\"{a}t des Saarlandes, Campus E2 9, 66041 Saarbr\"{u}cken, Germany}
\fntext[fn1]{now at: Institute for Applied Geophysics and Geothermal Energy, RWTH Aachen, Mathieustr. 10, 52074 Aachen, Germany}
\fntext[fn2]{now at: Department of Geosciences, University of Oslo, Postboks 1022 Blindern, 0315 Oslo, Norway}

\begin{abstract}
	Multi--color Stochastic Rotation Dynamics ($\MCSRD$) has been introduced by Inoue et al.~\cite{Inoue2004,Inoue2006} as a particle based simulation method to study the flow of emulsion droplets in non--wetting microchannels. In this work, we extend the multi--color method to also account for different wetting conditions. This is achieved by assigning the color information not only to fluid particles but also to virtual wall particles that are \oldtext{needed to apply}\newtext{required to enforce} proper no--slip boundary conditions. To \oldtext{reproduce the correct hydrodynamics}\newtext{extend the scope of the original}\oldtext{we reformulate the} $\MCSRD$ algorithm \newtext{to e.g. immiscible two--phase flow with viscosity contrast} \newtext{we implement an}\oldtext{that additionally conserves} angular momentum \newtext{conserving scheme} ($\MCSRDp$). We perform\oldtext{ed} extensive \newtext{benchmark} simulations to \newtext{show that a mono--phase $\MCSRD$ fluid exhibits bulk properties identical to a standard SRD fluid and that $\MCSRD$ fluids are applicable to a wide range of immiscible two--phase flows.}\oldtext{determine the dynamic shear viscosity and interfacial tension of the $\MCSRDp$ fluid. The interfacial tension was measured globally with the Young--Laplace test and locally from microscopic stress profiles and were in very good agreement. Viscosity and interfacial tension were validated by studying the deformation of a viscous drop in linear shear flow.} To quantify the adhesion of \oldtext{the}\newtext{a} $\MCSRDp$ fluid\oldtext{s} in contact to the walls we measure\oldtext{d} the apparent contact angle from sessile droplets in mechanical equilibrium. \newtext{For a further verification of our wettability implementation}\oldtext{Additionally} we compare the dewetting of a liquid film from a wetting stripe to experimental and numerical studies of interfacial morphologies on chemically structured surfaces.
\end{abstract}


\journal{Journal of Computational Physics}

\maketitle


\section{Introduction}

Capillarity dominated flows of immiscible fluids on the micro--scale are of central importance in many coating processes~\cite{Kistler1997,Bonn2009}, secondary oil recovery~\cite{Morrow1990,Sahimi2011} or the advancing field of microfluidics~\cite{Stone2004,Seemann2012}. Motion of fluid interfaces and their topological changes such as droplet pinch--off or coalescence are difficult to capture by finite element methods and become even more complex in the presence of rigid walls~\cite{Sui2014}. Capillary flows with wall contact depend crucially on wettability~\cite{Stone2004,Seemann2012} but \oldtext{there is a lack of numerical methods able to reliably capture these features}\newtext{the disproportionately high computational costs to capture the specific wall interactions in e.g. level--set or phase field models make it virtually impossible to study large scale systems}.

Over the last decades several particle based methods including dissipative particle dynamics (DPD)~\cite{Hoogerbrugge1992,Espanol1995}, Lattice Boltzmann (LB)~\cite{Shan1993,Chen1998,Succi2001} or multi particle collision dynamics (MPC)~\cite{Malevanets1999,Malevanets2000a,Kapral2008,Gompper2008} have been developed to study a wide range of soft condensed matter systems on the meso--scale. Particle based methods share the idea of a coarse graining procedure that lumps together the microscopic degrees of freedom of the fluid particles into larger macroscopic entities that, after suitable spatial and temporal averaging, display the fluid mechanical properties. The DPD method, closely related to Molecular dynamics (MD)~\cite{Haile1992} is still too detailed to provide an efficient Navier--Stokes solver and the integration of Newton's equation of motion comes with high numerical costs. Although widely used in computational sciences LB models have some limitations especially when enforcing certain boundary \newtext{conditions}. Because LB models consider particle populations with discrete velocities residing on a regular spatial lattice, embedded objects that are \newtext{of} irregular \newtext{shape} or off--lattice lead to further treatment of the fluid--solid interface by e.g. immersed boundary methods~\cite{Mittal2005,Kaoui2012}.

More recently the MPC method introduced by Malevanets and Kapral~\cite{Malevanets1999} has gained attraction in the field of computational fluid dynamics. It provides a robust method to obtain the correct transport of mass, momentum, and energy on the macro--scale. In their pioneering work on polymer solutions in meso--scale systems~\cite{Malevanets2000a} the authors coupled MPC to simulate the solvent and MD to study the solute dynamics. This hybrid approach has since been used to study equilibrium colloidal suspensions~\cite{Hecht2005,Padding2006,Goetze2007,Kapral2008,Gompper2008,Franosch2011} and polymer~\cite{Kapral2008,Gompper2008,Huang2010,Huang2013} solutions. An even more eminent relevance to real applications is the use of MPC to study systems out of equilibrium that are driven by flow including colloids~\cite{Allahyarov2002,Lamura2002,Padding2004,Goetze2010b,Goetze2011}, polymers~\cite{Huang2010,Ripoll2006,Canna2008,Frank2008}, liquid crystals~\cite{Lee2015} and fluid vesicles or blood cells~\cite{Noguchi2004,Noguchi2005,McWhirter2008}. Furthermore, MPC was also successfully applied to study bacteria~\cite{Reigh2012}, sperm cells~\cite{Elgeti2010} and swimmers and squirmers in general~\cite{Earl2007,Goetze2010,Theers2013,Theers2014}.

The method employed in this work belongs to a subset of MPC methods termed stochastic rotation dynamics (SRD). The name originates from the specific realization of momentum exchange between fluid particles during collisions. In all SRD variants the diffusive transport of momentum is achieved through a stochastic rotation of the relative velocities of the particles in a collision cell~\cite{Malevanets1999,Ihle2001,Ihle2003,Ihle2003a}. In the course of this work we will use the term SRD rather than MPC even though some general statements may refer to both types of methods.

\oldtext{In the past}\newtext{In recent years}, different SRD variants have been used to model phase separating binary and ternary fluid mixtures~\cite{Hashimoto2000,Sakai2002,Tuezel2007}. A modified SRD algorithm that accounts for an arbitrary number of fluid phases has been proposed by Inoue et al.~\cite{Inoue2004,Inoue2006,Inoue2008}. Inoue's multi--color algorithm ($\MCSRD$) employs a collision operator that actively maintains a segregation of particles with different colors.
Whilst the multi--color model accounts for phase immiscibility, the interaction of the fluids with the walls, or embedded objects with different \oldtext{contact angles}\newtext{wall affinities} was not yet addressed. To this end, we implemented an extension to the $\MCSRD$ scheme in order to account also for surface wettability. Especially for capillary dominated flows where the fluid--surface interaction \oldtext{becomes important}\newtext{is of central importance}~\cite{Stone2004,Seemann2012} this extension can be employed to study colloidal suspensions in immiscible fluid phases, porous media, micro-- or nanofluidics or other fields of soft condensed matter.

\newtext{The standard SRD method is a well established tool to study mono--phase fluids on the meso--scale and its properties have been thoroughly investigated by several authors over the last years\cite{Ihle2001,Ihle2003,Ihle2003a,Tuezel2003,Ihle2004,Ripoll2004,Ihle2005,Pooley2005,Tuezel2006,Goetze2007,Noguchi2008,Winkler2009,Huang2010a,Petersen2010,Whitmer2010,Hanot2013,Theers2015}. In their introductory work Inoue et al.~\cite{Inoue2004} only measured the surface tension qualitatively for a 2D droplet and showed that the Brownian motion of the center of mass of a droplet follows a Maxwell--Boltzmann distribution. What the $\MCSRD$ method has been lacking so far is the characterization of bulk fluid properties as well as the interaction of two immiscible phases, especially in three dimensions. To this end, and before introducing our wettability implementation, we perform a series of benchmark simulations to determine the relevant hydrodynamic properties of a $\MCSRD$ fluid. To the best of our knowledge this is the first time that such an extensive study is carried out for $\MCSRD$ fluids to verify the reliability of the method. To present a coherent description this}\oldtext{This} work is outlined as follows. \oldtext{In Sec.}\newtext{Section}~\ref{sec:method} \oldtext{we present}\newtext{introduces} the theoretical foundations of the methods employed in this work. In Sec.~\ref{ssec:viscosity} we determine the dynamic viscosity of \oldtext{the}\newtext{a} $\MCSRD$ fluid from local measurements of the shear rate and stress tensor in a linear shear flow. In Sec.~\ref{sec:twophase} we determine the interfacial tension between two immiscible fluid phases with three independent methods and verify the \newtext{beforehand determined} values with predictions for the deformation of a droplet in a linear shear flow. In Sec.~\ref{sec:wetting} we present our extension to the $\MCSRD$ scheme that accounts for varying surface wettability. We test our wetting implementation on homogeneously and heterogeneously wettable surfaces and compare the \oldtext{accompanied}\newtext{resulting} interfacial configurations with previous experiments and numerical studies.

\section{Model and Methods}
\label{sec:method}

In the following we briefly present the standard \newtext{SRD} algorithm \oldtext{based on the SRD method}(\ref{ssec:mpcmodel}) and an extension that respects angular momentum conservation (\ref{ssec:angmomentum}). After introducing the $\MCSRD$ algorithm of Inoue et al.~\cite{Inoue2004} in Sec.~\ref{ssec:multiphase} we present our implementation of stress measurements based on \oldtext{areal}\newtext{area--weighted} averages (\ref{ssec:stresscalc}). This allows us to localize very precisely the momentum transport inside a collision cell. The necessity of this approach is shown later \oldtext{on} in the course of this work (see Sec.~\ref{ssec:planar}).

\subsection{Stochastic Rotation Dynamics}
\label{ssec:mpcmodel}

Particle based simulation methods obtain the collective dynamics of the fluid phases from the motion of a large number $N$ of point particles $i$ of mass $m$ that can adopt continuous positions, $\mathbf{x}_{i}$, and velocities, $\mathbf{v}_{i}$, in three dimensional Euclidean space. The dynamics of the particles consist of a sequence of streaming and collision steps. During free streaming, particles move deterministically between time $t$ and $t+\Delta t$. New positions $\mathbf{x}_{i}(t+\Delta t)$ and velocities $\mathbf{v}^\prime_{i}(t+\Delta t)$ at the end of the streaming step are consequently given by
\begin{equation}
  \mathbf{x}_{i}(t+\Delta t)=\mathbf{x}_{i}(t) + \mathbf{v}_{i}(t)
  \Delta t + \frac{\mathbf{f}_{\rm ex}}{2m_{i}} \Delta t^2~,
  \label{eq:streaming1}
\end{equation}
and
\begin{equation}
  \mathbf{v}^\prime_{i}(t+\Delta t)=\mathbf{v}_{i}(t) +
  \frac{\mathbf{f}_{\rm ex}}{m_{i}} \Delta t~,
  \label{eq:streaming2}
\end{equation}
respectively, where a constant external force $\mathbf{f}_{\rm ex}$ acting on all particles is included. In our notation all pre--collisional quantities that may change during collision are marked with a prime. The corresponding post--collisional quantities (at the beginning of the next streaming step) are not primed. 

In order to introduce an interaction among particles, i.e.~an exchange of linear momentum, the particles are sorted into collision cells $\xi$ after each streaming step. In the present algorithm we use a cubic grid of uniform spacing $a$ where the number of particles per cell, $N_{\xi}$, fluctuates around an average value $\Navg$.

In every collision, which occurs instantaneously at time $t+\Delta t$ the velocities of the particles are decomposed into the center of mass velocity $\mathbf{u}_{\xi}$ of all particles belonging to cell $\xi$ and a remaining, fluctuational part $\widetilde{\mathbf{v}}_{i}^\prime=\mathbf{v}^\prime_{i}-\mathbf{u}_{\xi}$. In all variants of SRD, linear momentum between particles in a cell is exchanged through a rotation of the fluctuational velocity components. The particle velocity after the effective collision step is
\begin{equation}
  \mathbf{v}_{i}(t+\Delta t)=\mathbf{u}_{\xi}(t+\Delta t) +
  \boldsymbol{\Omega}_\xi\,\left\{\widetilde{\mathbf{v}}^\prime_{i}(t+\Delta
    t)\right\}~,
  \label{eq:collision1}
\end{equation}
where $\boldsymbol{\Omega}_\xi$ denotes a rotation around an axis $\mathbf{R}_\xi$ by an angle $\alpha$. In order to achieve molecular chaos, the unit vector $\mathbf{R}_\xi$ is randomly drawn from the surface of the three dimensional unit sphere for every cell $\xi$ and in every collision step. Most implementations of the mono--phase SRD algorithm employ a fixed rotation angle $\alpha$. Density correlations of the fluid particles which may occur for small mean free paths $\MFP = \Delta t\sqrt{k_{B}T/m} \ll a$ are avoided through a shift of the collision grid before each collision step. The Cartesian components of the random shift $\boldsymbol{\zeta}$ are drawn uniformly from the interval $[-a/2,a/2]$~\cite{Ihle2001}.

As the particle collisions in standard SRD obey conservation of mass, linear momentum and energy we observe both diffusive and advective transport of the conserved quantities on length scales larger than the grid spacing $a$. The evolution of spatial and temporal averages of hydrodynamic quantities like velocity $\mathbf{u}$ and mass density $\rho=nm$, with the number density $n=N_{\xi}/V_{\xi}$ in the SRD fluid conform to a continuum description by the Navier--Stokes equation \cite{Malevanets1999}. It was shown that the detailed balance condition for the SRD collisions is satisfied and that therefore an H theorem exists \cite{Malevanets1999,Ihle2003,Ihle2003a}. The equation of state of a SRD fluid is identical to the equation of state of an ideal gas. However, recently it was shown that SRD fluids exhibit a nonzero bulk viscosity which seems to contradict the ideal gas assumption \cite{Theers2015}. As the bulk viscosity is determined by the particle collisions the ideal gas limit is reached for large collision time steps. This means a collisional dominated system (short mean free path $\MFP$, non--zero bulk viscosity) is more \emph{liquid--like} whereas a kinetic dominated system (large mean free path $\MFP$, vanishing bulk viscosity) is more \emph{gas--like}~\cite{Theers2015}.

Several authors derived explicit expressions for the transport coefficients for a mono--phase SRD fluid based on the Green--Kubo formalism~ \cite{Ihle2003,Ihle2003a,Tuezel2003,Ihle2005}. Alternatively to this equilibrium approach the transport coefficients can also be measured out of equilibrium, like e.g. in shear flow. This was first shown in Ref.~\cite{Ihle2001} and later extended in Refs.~\cite{Kikuchi2003,Pooley2005}. For a comprehensive summary, we refer to the overview given in Ref.~\cite{Gompper2008} and the references therein.

In general, SRD fluids are Newtonian with a kinematic viscosity in three dimensions as given by e.g.~\cite{Tuezel2003,Ihle2005}
\begin{equation}
\nu\equiv\frac{\eta}{\rho} =\frac{k_{B}T\Delta t}{2m} \frac{5\Navg}{(\Navg-1+e^{-\Navg})[2-\cos{\alpha}-\cos{2\alpha}]} + \frac{a^2}{\Delta t}\frac{\Navg-1+e^{-\Navg}}{18\Navg}\left(1-\cos\alpha\right)~,
\label{eq:kinvisc}
\end{equation}
where $\eta$ denotes the dynamic viscosity and $k_{B}T$ is the thermal energy scale.

When solid walls are introduced in a SRD system the fluid dynamic boundary conditions need to be considered. In order to guarantee a no--slip boundary condition for the average fluid velocities, a generalization of the bounce--back rule for partially filled cells is employed~\cite{Lamura2001}. In the na\"ive formulation of the bounce--back rule, particles travel back into the direction of their incidence after having collided with the solid boundaries during free streaming. Because the position of the solid boundaries relative to the coarse--graining grid changes between every inter--particle collision step due to the random grid shift, it is necessary to add a virtual phase resting inside the walls to match the bulk particle density in the underfilled cells. These virtual wall--particles participate in the collisions and are generated before and removed after every inter--particle collision step, and guarantee a no--slip boundary condition at the walls~\cite{Lamura2001}. If, however, partial slip is desired at the solid walls it is possible to tune the local particle density inside the walls \cite{Hanot2013} or choose an alternate reflection mechanism \cite{Whitmer2010} and vary therewith the local slip length. Note, that in general the method of virtual particles is only needed for systems where the viscosity is dominated by the collisional contribution e.g. short mean free paths $\MFP$. When the mean free path is larger than $\MFP \gtrsim 0.6\:a$ and ergo the viscosity is dominated by the kinetic contribution, the no--slip boundary condition is readily achieved provided that the bounce--back rule is applied~\cite{Bolintineanu2012}. 

Unless explicitly defined we set grid size $a$, collisional time step $\Delta t$, particle mass $m$ and the Boltzmann constant $k_{B}$ to unity. If these rescaled units are used, the mean free path of the particles $\MFP$ depends only on temperature $T$.

\subsection{Angular momentum conservation and thermostatting}
\label{ssec:angmomentum}

In their basic formulation \newtext{MPC/}SRD algorithms do not conserve angular momentum \cite{Malevanets1999,Goetze2007,Goetze2010} but can be \newtext{easily} extended to do so \cite{Noguchi2008}. Generally, when a mono--phase system is considered, the lack of angular momentum conservation only modifies the viscosity of the fluid \cite{Goetze2007}. However, following Ref.~\cite{Goetze2007} there are several cases where angular momentum conservation is essential:
\begin{enumerate}[1.]
	\item the boundary conditions on walls are given by forces including torques (circular Couette flow)
	\item finite--sized objects that rotate in fluids by hydrodynamic stress (colloidal and polymer suspensions)
	\item fluids with different viscosities are in contact.
\end{enumerate}
Especially the last point becomes important when studying multi--phase or microfluidic systems where the fluids involved often have a certain viscosity contrast. Thus, we like to point out that \newtext{especially in $\MCSRD$} vorticity is a hydrodynamical degree of freedom \oldtext{and}\newtext{that} needs to be locally conserved during collisions. \newtext{Because in standard SRD implementations angular momentum conservation depends on the predefined collision angle $\alpha$ another approach has to be used in $\MCSRD$ where the collision angle $\alpha$ is calculated individually for every collision cell. We employ a straight forward protocol where}\oldtext{Therefore,} the angular momentum in every collision cell is first calculated and then subtracted from the individual particle velocities \newtext{before the actual collision}. This procedure ensures that the fluctuational velocities subject to the collision operator are effectively irrotational.

To enforce local conservation of angular momentum the following steps are introduced into the collision operation from eqn.~\eqref{eq:collision1} at time $t+\Delta t$. The total angular momentum of a collision cell $\mathbf{L}^{\prime}_{\xi}$ is calculated by
\begin{equation}
\mathbf{L}^{\prime}_{\xi} = \sum_{i=1}^{N_{\xi}}{\widetilde{\mathbf{x}}_{i} \times m_{i}\mathbf{v}^{\prime}_{i}}~,
\label{eq:AM1}
\end{equation}
where $\widetilde{\mathbf{x}}_{i}$ is the relative position of particle $i$ to the center of mass of the collision cell $\xi$. Additionally the moment of inertia tensor $\mathbf{I}_{\xi}$ is calculated by
\begin{equation}
{I}_{\xi}^{\alpha\beta} = \sum_{i=1}^{N_{\xi}}m_{i}\left(r_{i}^{2}\delta_{\alpha\beta} - \widetilde{x}_{i\alpha}\widetilde{x}_{i\beta}\right)~,
\label{eq:AM2}
\end{equation}
with $\mathbf{r}_{i}$ the vector pointing from the center of mass to particle $i$ and $\delta_{\alpha\beta}$ the Kronecker symbol. Then the vorticity $\boldsymbol{\omega}^{\prime}_{\xi}$ is given by
\begin{equation}
\boldsymbol{\omega}^{\prime}_{\xi} = 2\mathbf{I}_{\xi}^{-1}\mathbf{L}^{\prime}_{\xi}~,
\label{eq:AM3}
\end{equation}
which is twice the angular velocity. The contribution of the angular momentum of the cell on particle $i$ is then calculated by
\begin{equation}
\widehat{\mathbf{v}}^{\prime}_{i}=\frac{1}{2}\boldsymbol{\omega}^{\prime}_{\xi} \times \widetilde{\mathbf{x}}_{i}~.
\label{eq:AM4}
\end{equation}
Consequently, when considering the velocity from eqn.~\eqref{eq:AM4} the actual relative velocity of particle $i$ is
\begin{equation}
\widetilde{\mathbf{v}}^{\prime}_{i}=\mathbf{v}^{\prime}_{i}-\mathbf{u}_{\xi}-\widehat{\mathbf{v}}^{\prime}_{i}~,
\label{eq:AM5}
\end{equation}
so that the fluctuational velocities $\widetilde{\mathbf{v}}^{\prime}_{i}$ in a collision cell are irrotational. The fluctuational velocities $\widetilde{\mathbf{v}}^{\prime}_{i}$ from eqn.~\eqref{eq:AM5} are then subjected to the standard SRD collision. As the collision operator itself generates angular momentum on the post--collisional fluctuational velocities $\widetilde{\mathbf{v}}_{i}$ this additional vorticity needs to be removed. Therefore, the post--collisional angular momentum $\mathbf{L}_{\xi}$, vorticity $\boldsymbol{\omega}_{\xi}$ and the correction to the fluctuational velocity $\widehat{\mathbf{v}}_{i}$ are calculated similar to eqns.~(\ref{eq:AM1},~\ref{eq:AM3},~\ref{eq:AM4}) by
\begin{equation}
\mathbf{L}_{\xi} = \sum_{i=1}^{N_{\xi}}{\widetilde{\mathbf{x}}_{i} \times m_{i}\widetilde{\mathbf{v}}_{i}}~,
\label{eq:AM6}
\end{equation}
\begin{equation}
\boldsymbol{\omega}_{\xi} = 2\mathbf{I}_{\xi}^{-1}\mathbf{L}_{\xi}~,
\label{eq:AM7}
\end{equation}
and
\begin{equation}
\widehat{\mathbf{v}}_{i} = \frac{1}{2}\boldsymbol{\omega}_{\xi} \times \widetilde{\mathbf{x}}_{i}~.
\label{eq:AM8}
\end{equation}
After the collision procedure the new particle velocities are
\begin{equation}
\mathbf{v}_{i} = \mathbf{u}_{\xi} + \widetilde{\mathbf{v}}_{i} - \widehat{\mathbf{v}}_{i} + \widehat{\mathbf{v}}^{\prime}_{i}~,
\label{eq:AM9}
\end{equation}
where the contribution from the collision operator $\widehat{\mathbf{v}}_{i}$ to the angular momentum is removed and the pre--collisional contribution $\widehat{\mathbf{v}}^{\prime}_{i}$ is added back. In this way the pre--collisional angular momentum $\mathbf{L}^{\prime}_{\xi}$ is restored. The method employed here effectively conserves angular momentum with an additional computational overhead of $\approx 20\%$ tested up to a total amount of $10^7$ particles per system. This increase in computational time is comparable to values reported for angular momentum conserving MPC methods with Anderson thermostat (MPC--AT$+\alpha$)~\cite{Gompper2008}. In the course of this work we will distinguish between the angular momentum conserving ($\MCSRDp$) and not conserving ($\MCSRDm$) case, respectively.

In any non--equilibrium MPC/SRD simulation with external driving of the particles the control of the system temperature is essential. Injection of work into the MPC/SRD fluid and dissipation through viscous heating may occur through external forces or by imposed motion of the walls confining the particles. A standard method to enforce a constant temperature in a MPC fluid is to implement an Anderson thermostat (MPC--AT)~\cite{Allahyarov2002,Goetze2007,Noguchi2007}. Instead of rotating the relative velocities in a collision cell, new relative velocities are generated at each time step from a Maxwell Boltzmann distribution with zero mean and temperature dependent standard deviation. Obviously, an Anderson thermostat is not applicable to SRD based algorithms because during collisions relative velocities are rotated rather than newly generated.

In continuation with the standard SRD method and due to its simplicity in implementation we use a profile--unbiased local thermostat (PUT) in this work. This ensures control of the thermal fluctuations while keeping unaffected the macroscopic velocity field~\cite{Evans1986,Noguchi2007,Huang2010a}. In a recent study PUT was compared to a Maxwell--Boltzmann scaling (MBS) method as introduced in Ref.~\cite{Huang2010a} and in regard to the measured fluid viscosities was found to perform equally well~\cite{Huang2015}. The fluctuational velocities of each cell $\xi$ are rescaled after each collision step as $\widetilde{\mathbf{v}}_{i} \rightarrow \lambda_{\xi}\widetilde{\mathbf{v}}_{i}$, where the correction factor is calculated by
\begin{equation}
\lambda_{\xi} = \sqrt{\frac{3(N_{\xi}-X)T}{\sum_{i=1}^{N_{\xi}}{m_{i}(\widetilde{\mathbf{v}}_{i}-\mathbf{u}_{\xi})^{2}}}}~.
\label{eq:thermo1}
\end{equation}
The term $3(N_{\xi}-X)$ accounts for the particles spatial degrees of freedom (DOF) in three dimensions minus the DOF of the center of mass of the collision cell. If angular momentum is not conserved then $X=1$ (only translational DOF). In case of angular momentum conservation $X=2$ (translational and rotational DOF) and $\widetilde{\mathbf{v}}_{i}$ in eqn.~\eqref{eq:thermo1} has to be replaced by  $\widetilde{\mathbf{v}}_{i} - \widehat{\mathbf{v}}_{i}$ to account for the additional rotational components.

\subsection{Multi--phase implementation}
\label{ssec:multiphase}

The $\MCSRD$ algorithm of Inoue et al.~\cite{Inoue2004} utilizes a modified collision operator that actively creates a repulsive interaction between different particles species, but still allows for a diffusive momentum exchange between particles in the homogeneous phases. Different particle species (phases) are introduced through colors $c = 1,\dots,N_{p}$ assigned to each individual particle. At each collision step the color flux
\begin{equation}
  \mathbf{q}_{c}(\xi) =
  \sum_{i=1}^{N_{\xi}}\left({\mathbf{v}^{\prime}_{i}-\mathbf{u}_{\xi}}\right)\,\delta_{c\,c_{i}},
  \label{eq:flux1}
\end{equation}
is computed while color--gradients $\boldsymbol{\nabla} n_{c}(\xi)$ are estimated for each color $c$ in each cell $\xi$ from the number of particles of the same color in the next--nearest neighboring cells. Then, after selecting a random rotation axis $\mathbf{R}_{\xi}$ in each cell, the rotation angle $\alpha_{\xi}$ is constructed such that, the color action
\begin{equation}
  S_{\xi} = \sum_{c,c^\prime}^{N_{p}}
  \,\kappa_{c\,c^\prime}\,{\mathbf{q}_{c}(\xi) \cdot \boldsymbol{\nabla}n_{c^\prime}(\xi)},
  \label{eq:potential}
\end{equation}
in cell $\xi$ is maximized. The symmetric interaction matrix $\kappa_{c\,c^\prime}$ weights the relative tendency of colored particles to repel or attract each other. The necessary condition for a maximum
\begin{equation}
  \left.\frac{\partial S_{\xi}(\alpha)}{\partial
    \alpha}\right|_{\alpha=\alpha_{\xi}} = 0~,
  \label{eq:minu}
\end{equation}
has two solutions $\alpha_{\xi}^{\rm +}$ and $\alpha_{\xi}^{\rm -}$ in $[-\pi,\pi]$. Given the color gradients and the color fluxes the rotation angle $\alpha_\xi$ that satisfies condition eqn.~\eqref{eq:minu} can be computed from expression 
\begin{equation}
  \tan{\alpha_{\xi}} = \frac{\mathbf{R}_{\xi}\cdot
    \sum_{c}^{N_{p}}\,{\mathbf{q}_{c}(\xi)\times
      \mathbf{F}_{c}(\xi)}}{\sum_{c}^{N_{p}}{\mathbf{q}_{c}(\xi)\cdot
      \mathbf{F}_{c}(\xi)}}~,
	\label{eq:theta3}
\end{equation}
employing the weighted color gradient
\begin{equation}
  \mathbf{F}_{c}(\xi)=
  \sum_{c^\prime}^{N_{p}}\,{\kappa_{c\,c^\prime}\,\boldsymbol{\nabla}n_{c^\prime}}
  \label{eq:grad}
\end{equation}
is used. In order to guarantee that the color action is in a maximum after the rotation the actual rotation \newtext{(collision)} angle $\alpha_{\xi}$ must be chosen from the two solutions of eqn.~\eqref{eq:theta3} according to the condition
\begin{equation}
  \alpha_{\xi} =
  \begin{cases}
    \alpha_{\xi}^{\rm +} & {\rm if}\quad S_{\xi}(\alpha_{\xi}^{\rm +})>0 \\ 
    \alpha_{\xi}^{\rm -} & {\rm if}\quad S_{\xi}(\alpha_{\xi}^{\rm +})<0~.
  \end{cases}
  \label{eq:theta3a}
\end{equation}
Phase segregation between particle species $c \neq c^\prime$ is achieved by a negative sign $\kappa_{c\,c^\prime}<0$, as the particles of color $c$ are forced to move to regions where the concentration of particles with color $c^\prime$ is low. The opposite sign $\kappa_{c\,c^\prime}>0$ leads to mixing of particles of color $c$ and $c^\prime$. Diagonal entries are set to unity.

\subsection{Stress measurement}
\label{ssec:stresscalc}

In its most fundamental definition, stress is a flux of linear momentum and is described in three dimensions by a tensor $\boldsymbol{\sigma}$ of rank two with nine independent components. Transport of linear momentum in a fluid proceeds either through collective motion of particles during streaming or by exchange of linear momentum between particles during collisions. Referring to these two modes of momentum transport the total stress $\boldsymbol{\sigma}$ can be split into a kinetic contribution $\boldsymbol{\sigma}^{\mathrm{(kin)}}$ and a collisional contribution $\boldsymbol{\sigma}^{\mathrm{(col)}}$, so that
\begin{equation}
\boldsymbol{\sigma} = \boldsymbol{\sigma}^{\mathrm{(kin)}} + \boldsymbol{\sigma}^{\mathrm{(col)}},
\label{eq:stress}
\end{equation}
holds. Despite its definition as a flux of momentum, stress is measured in most particle based simulations by volume averaging of suitable expressions of particle positions and velocities. Global averages of the stress tensor in the absence of external forces can be obtained from the virial theorem (see e.g.~\cite{Clausius1870,Catlow1990}). In Cartesian components the stress average then reads
\begin{equation}
\left\langle\sigma_{\alpha\beta}\right\rangle=\frac{1}{V}
\sum_{i=1}^{N}m_{i}\left\langle v_{i\alpha}{v}_{i\beta}\right\rangle -
\frac{1}{V\Delta t}\sum_{i=1}^{N}m_{i}\left\langle\left(v_{i\alpha}-v^{\prime}_{i\alpha}\right){r}_{i\beta}\right\rangle,
\label{eq:virial}
\end{equation}
where $V$ is the volume of the simulation domain, $N$ is the total number of particles in the system and $\mathbf{r}$ is the position of particle $i$ inside the system, respectively. Angular brackets $\langle\dots\rangle$ indicate time averages of the enclosed variable(s). The first term in eqn.~\eqref{eq:virial} refers to the kinetic contribution $\boldsymbol{\sigma}^{\mathrm{(kin)}}$ whereas the second term refers to the collisional contribution $\boldsymbol{\sigma}^{\mathrm{(col)}}$, respectively. Expressions analogous to eqn.~\eqref{eq:virial} have been used before to measure the stress tensor in mono--phase MPC/SRD fluids~\cite{Kikuchi2003,Noguchi2008,Winkler2009}.

\begin{figure}
	\includegraphics[width=0.98\columnwidth]{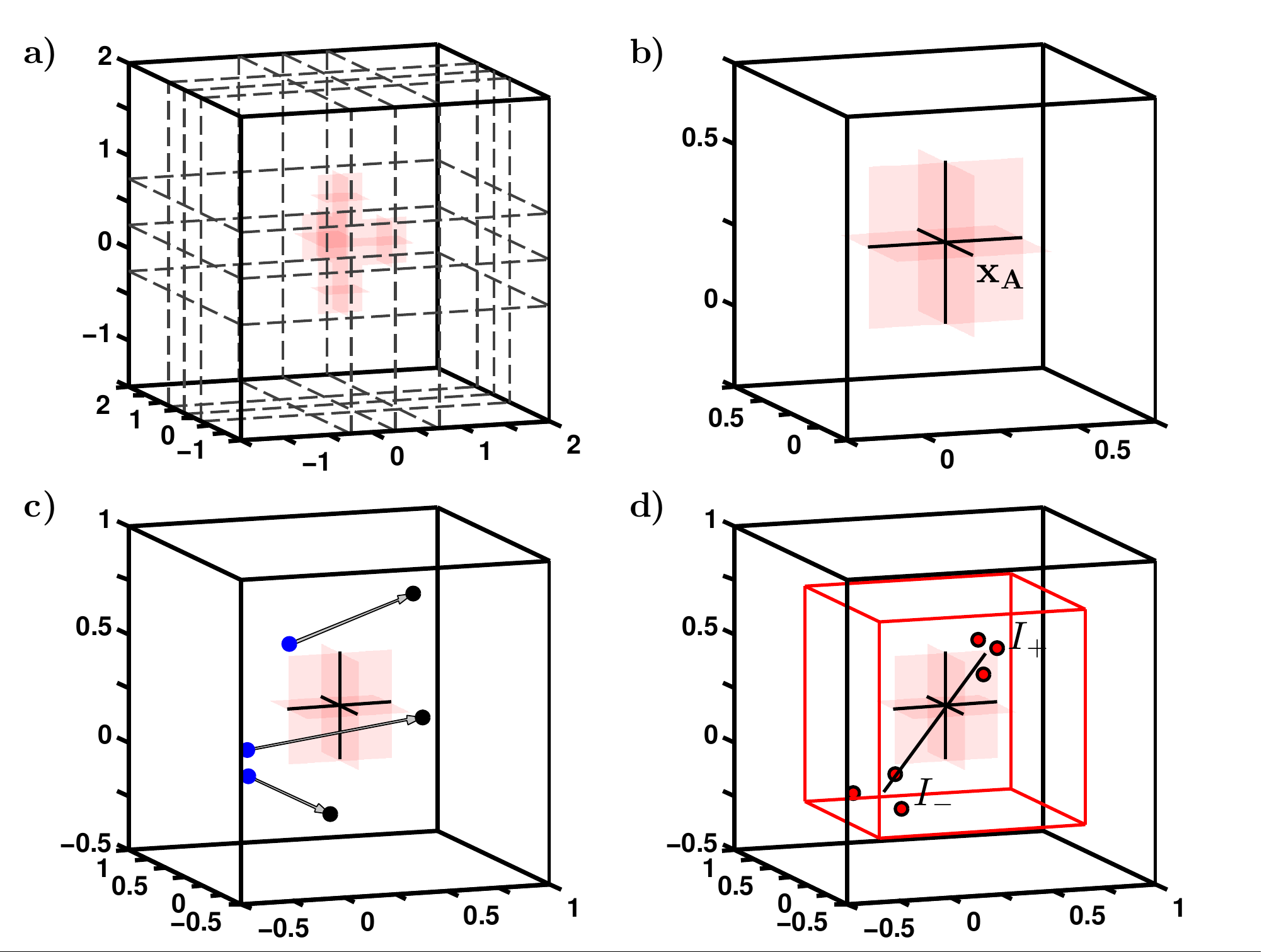}
	\caption{\small(color online) Measurement of the stress tensor components employing \oldtext{areal}\newtext{area--weighted} averages of momentum flux; \textbf{(a)} control planes for each spatial dimension with distance $d=a/2$ create a set of collocation points $\mathbf{x}_A$ and corresponding control surfaces $\mathcal{A}$ (red); \textbf{(b)} magnification of a single collocation point $\mathbf{x}_A$, the area of a single control surface is $A_{\mathcal{A}}=d^2=a^2/4$; \textbf{(c)} gray arrows indicate the particle motion during free streaming and their contribution to $\boldsymbol{\sigma}^{\mathrm{(kin)}}$ when crossing a control surface; \textbf{(d)} the red frame represents a collision cell and the black line connects the two center of mass for particles belonging either to $I_+$ or $I_-$, only the momentum exchanged between particles within the same collision cell contributes to the collisional stress tensor $\boldsymbol{\sigma}^{\mathrm{(col)}}$.}
	\label{fig:stressH}
\end{figure}

The definition of stress as momentum flux provides an alternative approach to volume averaging methods. In the following part we describe the measurement of the stress tensor by \oldtext{areal}\newtext{area--weighted} averages as outlined e.g. for systems with many--body interactions by Refs.~\cite{Todd1995,Heinz2005}. In this way global and local contributions to the stress tensor can be determined. To compute the local flux of linear momentum in the $\MCSRD$ fluid, the simulation domain is subdivided by control planes crossing the system in all three dimensions. In particular, we consider three stacks of planes normal to the direction of the three unit vectors $\mathbf{e}_x$, $\mathbf{e}_y$, and $\mathbf{e}_z$ of the Cartesian coordinate system. The control planes partition the simulation box into small cubes. The distance $d$ between every pair of neighboring planes is uniform and equals a certain fraction of the size of the collision grid $a$. Each stress plane is tiled into an array of small squares $\mathcal{A}$ with the lateral dimensions $d$. Each small square $\mathcal{A}$ is oriented normal to $\mathbf{e}_\alpha$ and centered around the intersection points $\mathbf{x}_A$ of the three squares $\mathcal{A}(\mathbf{x}_A,\alpha)$ into the three directions $\mathbf{e}_\alpha$ with $\alpha \in \{x,y,z\}$. All stress components measured in the $\MCSRD$ simulations are collocated at the intersection points $\mathbf{x}_A$.

Figure~\ref{fig:stressH} illustrates the stress measurement in our $\MCSRD$ simulations. Panel (a) shows a small fraction of a three--dimensional system that is subdivided by three control planes for each spatial dimension and therewith forming a regular lattice of points $\mathbf{x}_A$ with corresponding small control surfaces $\mathcal{A}$ (red planes). In the example shown in Fig.~\ref{fig:stressH}, we chose the resolution of the \emph{stress grid} to be two times larger than the grid size $a$ so that the distance between the individual stress planes is $d=a/2$ and, hence, the area of a single sub--plane in three dimensions is $A_{\mathcal{A}}=d^2=a^2/4$. As an example, panel (b) shows the magnified region around a single point $\mathbf{x}_A$. In panels (c) and (d) the measurement of the kinetic and collisional contribution is exemplified.

During streaming (Fig.~\ref{fig:stressH}c) particles move from the pre--streaming (blue) to the post--streaming position (black), respectively. The local momentum flux accounts for all particles with index $i\in I(\alpha,\mathbf{x}_A)$ that cross the small control surface $\mathcal{A}(\mathbf{x}_A,\alpha)$ during the streaming step between time $t$ and $t+\Delta t$. The local contribution to the kinetic part of the stress tensor $\boldsymbol{\sigma}^{\mathrm{(kin)}}$ that is collocated at the intersection point $\mathbf{x}_{A}$ is given by
\begin{equation}
\sigma_{\alpha\beta}^{\mathrm{(kin)}}(\mathbf{x}_A) = \frac{1}{A\,\Delta t}\sum_{i \in I(\alpha,\mathbf{x}_A)}
\,m_i v_{i\beta}\,\sgn{\left(x_{i\alpha}(t+\Delta t)-x_{A\alpha}(t)\right)}~.
\label{eq:mopkin}
\end{equation}
Fluid particles that cross several control planes during streaming contribute to the momentum flux at more than one points in the stress grid. The signum function accounts for the direction in which particles $i \in I(\mathbf{x}_A,\alpha)$ have crossed the area $\mathcal{A}(\mathbf{x}_A,\alpha)$.

To obtain the collisional contribution (Fig.~\ref{fig:stressH}d) of momentum flux into the direction $\alpha$ in a collocation point $\mathbf{x}_A$, we first need to identify all collision cells $\xi \in C(\mathbf{x}_A,\alpha)$ that intersect with the small surface $\mathcal{A}(\mathbf{x}_A,\alpha)$. Because a random shift $\boldsymbol{\zeta}$ is applied to the collision cells prior to every collision step, the set $C(\mathbf{x}_A,\alpha)$ of potentially contributing cells may differ between subsequent steps. We partition the set of fluid particles in cells $\xi \in C(\mathbf{x}_A,\alpha)$ into particles $i\in I_+(\xi,\bf{x}_A,\alpha)$ with coordinate $x_{i\alpha}>x_{A\alpha}$ and a set of particles $i\in I_-(\xi,\mathbf{x}_A,\alpha)$ with $x_{i\alpha}\leq x_{A\alpha}$. Because the total momentum in a collision cell is conserved, the amount of momentum that is added to the particles $i\in I_-(\xi,\bf{x}_A,\alpha)$ is subtracted from the total momentum of particles $i\in I_+(\xi,\bf{x}_A,\alpha)$, and it is sufficient to consider the change of 
momentum only for particles $i \in I_+(\xi,\bf{x}_A,\alpha)$.

To decide at which control surface $\mathcal{A}(\mathbf{x}_A,\alpha)$ the linear momentum of fluid particles in the subsets is exchanged, we compute the center of mass of particles in $I_+(\xi,\mathbf{x}_A,\alpha)$ and $I_-(\xi,\mathbf{x}_A,\alpha)$. The function $\delta(\xi,\mathbf{x}_A,\alpha)$ equals one if the line connecting both center of mass crosses $\mathcal{A}(\mathbf{x}_A,\alpha)$ and is zero else. Using the collocation according to the center of mass, the total contribution of the collisional momentum flux to the stress in point $\mathbf{x}_A$ is then given by 
\begin{equation}
\sigma_{\alpha\beta}^{\mathrm{(col)}}(\mathbf{x}_A)= \frac{1}{A\,\Delta t}
\sum_{\xi\in C(\mathbf{x}_A,\alpha)}\sum_{i\in I_+(\xi,\mathbf{x}_A,\alpha)}\,\delta(\xi,\mathbf{x}_A,\alpha)\,m_i(v_{i\beta}-
v^\prime_{i\beta})~,
\label{eq:mopcol}
\end{equation}
with the pre-- and post--collisional velocity components $v^\prime_{i\beta}$ and $v_{i\beta}$. Special care has to be taken with the sign convention of the stress.

The \oldtext{areal}\newtext{area--weighted} averaging procedure can be used to measure the local stress field inside the simulation domain for an arbitrarily high resolution. It can be applied to the standard SRD collision operator introduced in Sec.~\ref{ssec:mpcmodel} as well as to the multi--color collision operator presented in Sec.~\ref{ssec:multiphase}. Computation of the collisional stress contribution effectively \emph{localizes} the momentum exchange to a regular lattice of points $\mathbf{x}_A$. It has been emphasized in a number of works (e.g.~\cite{Irving1950,Schofield1982,Todd1995}) that the freedom to localize the exchange of momentum on the control surface represents a gauge freedom for the stress field. The physical observable is the local force on a fluid element which must be independent on the gauge of the stress field.

\section{Dynamic viscosity of a mono--phase system}
\label{ssec:viscosity}

Before we present the results for multi--phase systems we will focus on the dynamic viscosity $\eta$ of a mono--phase SRD fluid subject to the $\MCSRD$ collision operator introduced in Sec.~\ref{ssec:multiphase}. Stress measurements according to \oldtext{areal}\newtext{area--weighted} averaging methods presented in Sec.~\ref{ssec:stresscalc} provide us with the possibility to compare differences in the single--phase properties of a fluid between the $\MCSRD$ algorithm and the standard SRD algorithm. Transport coefficients for the latter, including kinematic viscosity eqn.~\eqref{eq:kinvisc} are available in closed form expressions derived from the Green--Kubo relations \cite{Ihle2003,Ihle2003a} for systems in thermal equilibrium.

\begin{figure}
	\centering
	\includegraphics[width=0.49\columnwidth]{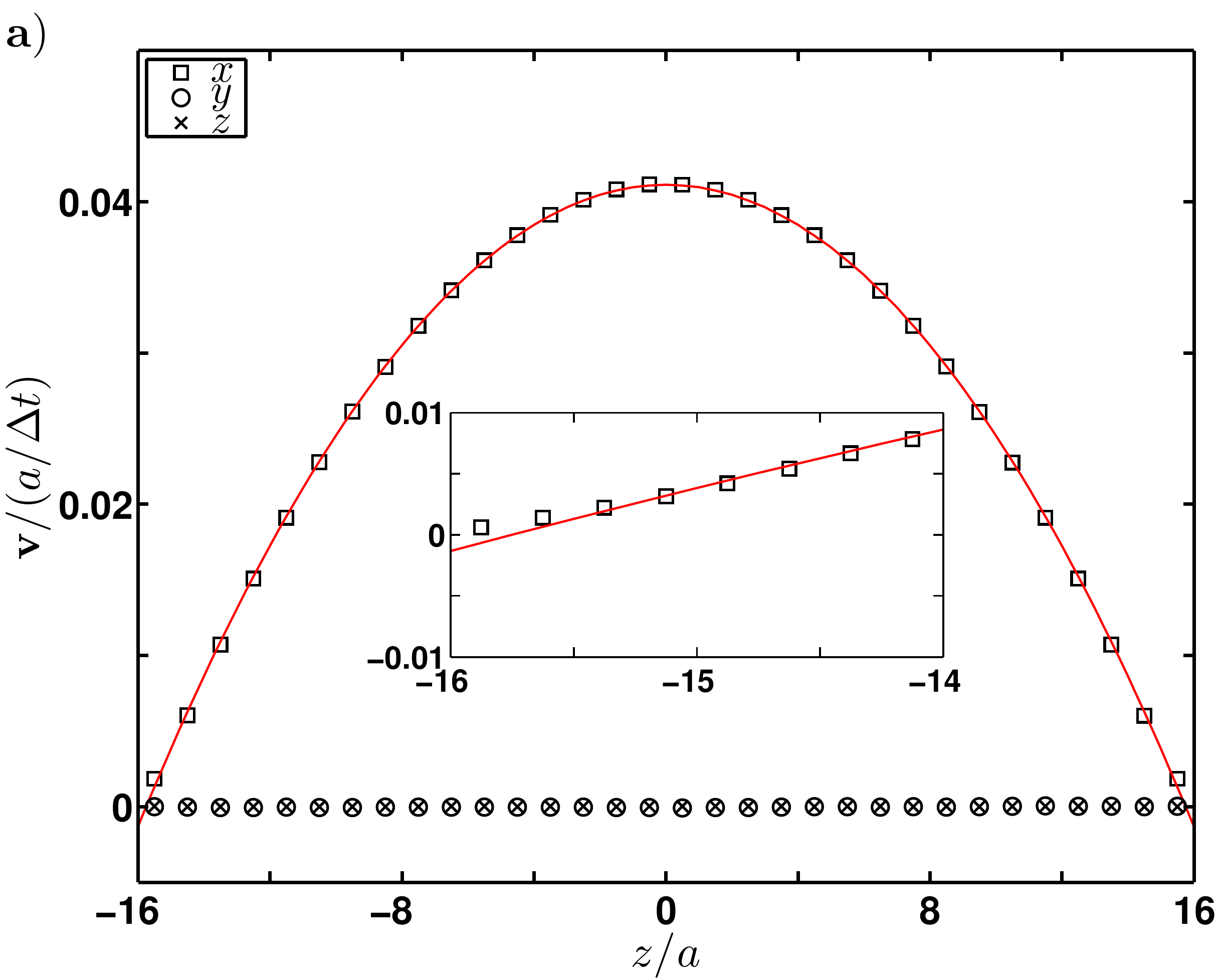}\hfill
	\includegraphics[width=0.49\columnwidth]{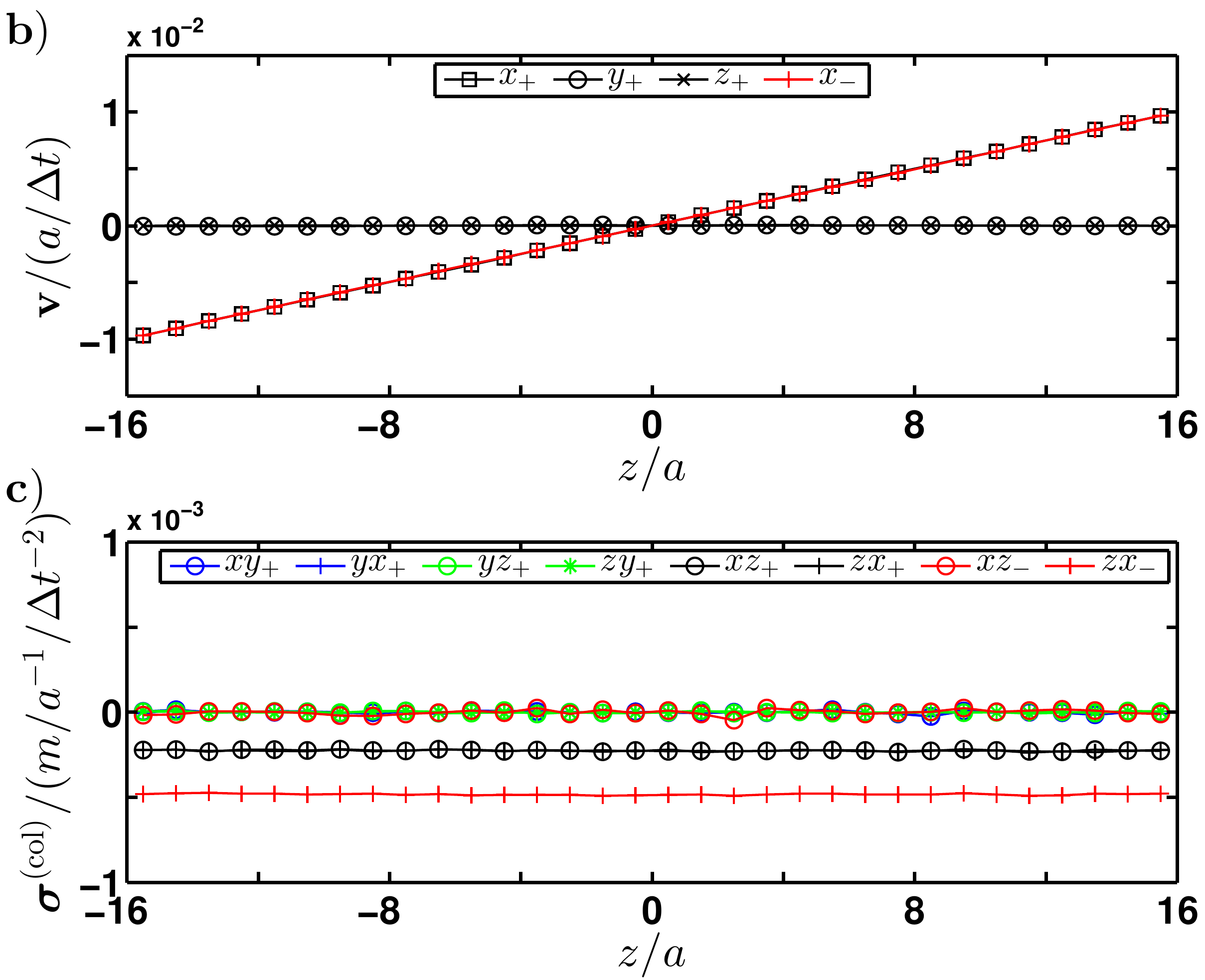}	
	\caption{(color online) (\textbf{a}) velocity profile for Poiseuille flow and corresponding parabolic fit, the spatial sampling rate inside the inset is higher by a factor of 4; (\textbf{b}) velocity profiles for a $\MCSRDp$ (black) and $\MCSRDm$ (red) fluid under linear shear; (\textbf{c}) off--diagonal elements of the collisional stress tensor $\boldsymbol{\sigma}^{\mathrm{(col)}}$ of the two systems from (b).}
	\label{fig:flow}
\end{figure}

\subsection{Poiseuille flow}

A straightforward approach to numerically determine the dynamic viscosity without measurement of the stress tensor is to confine the fluid between two parallel walls with a no--slip boundary condition at $z=\pm L/2$ and to apply a constant external force $\mathbf{f}_{ex}=f_{ex} \mathbf{e}_x$ to the fluid particles. Once a stationary state is reached, the average velocity profile $v_x(z)$ of the fluid particles follows a parabola $v_x(z) = C\,z(z-L)^2$ whose opening $C=f_{ex}/2\eta$ is determined by the dynamic viscosity $\eta$, see e.g. Ref.~\cite{Allahyarov2002}. Figure~\ref{fig:flow}a illustrates the measured Poiseuille flow in a cubic simulation box with dimensions $L^3=32^3$ and periodic boundary conditions in $x$-- and $y$--direction. A simple bounce--back BC is applied to the fluid particles at the walls $z=\pm L/2$ while the magnitude of the external force is $f_{ex}=10^{-5}$ in eqn.~\eqref{eq:streaming1}. To avoid viscous heating we apply a PUT thermostat as described in Sec.~\ref{ssec:angmomentum}. The average particle number is $\Navg=15$ and the temperature of the system is set to $T=5\times 10^{-3}$ with a corresponding mean free path $\MFP\approx 7.1\times 10^{-2}$. To enforce a proper no--slip BC on the walls we employ virtual fluid particles \cite{Lamura2001,Bolintineanu2012} in addition to the bounce--back rule, as described in Sec.~\ref{ssec:mpcmodel}. The velocity profile was averaged over $5 \times 10^4$ time steps after saturation to the stationary flow. The red curve in Fig.~\ref{fig:flow}a is the parabolic fit to the $x$--component of the velocity (open black squares).

At first glance the velocity profile displays the expected parabolic shape. A close inspection of the region close to the walls shown in the inset of Fig.~\ref{fig:flow}a reveals a negative apparent slip for this particular choice of control parameters. In accord with the bounce--back rule, we find that the average tangential velocity at the walls $z=\pm L/2$ tends to zero. The bending of the velocity profile from a negative second derivative $\partial^2_{zz} v_x$ away to positive values at distances below a lattice unit $a$ stems from a spatial variation of viscosity. An enhanced transport of tangential momentum between the particles in the bulk fluid and the walls in presence of virtual wall particles is caused by the random shift of the collision cells. This and similar effects were already discussed in literature and different ways have been proposed for a correct treatment of the wall cells, e.g.~Poisson distributed particle densities inside the walls \cite{Winkler2009}. Another approach was suggested by Ref.~\cite{Huang2015} where the authors assign finite (negative) velocities to the virtual wall particles depending on their position inside the wall. This led to a zero fluid velocity at the wall surface and viscosities close to the theoretical predictions.

Another source of error that may arise when determining the fluid viscosity from a parabolic profile is the finite size of the simulation box and employed collision cells, respectively. It may be possible that the transport of momentum is not only related to the local velocity gradient but also contains contributions of higher order derivations. Therefore, we repeated the Poiseuille flow experiment with a simulation box with dimensions $L^3=64^3$ but still the bending of the velocity profile close to the wall was noticeable. It has been addressed before that systems driven by Poiseuille flow are also very sensitive to the applied thermostat and that the derived viscosities can differ significantly (see e.g.~\cite{Whitmer2010,Bolintineanu2012,Huang2015}). Hence, we decided to use linear shear flow experiments for measuring the dynamic viscosity which exhibits linear velocity profiles and no dependencies on higher order derivations.

\subsection{Linear shear flow}

Local stress measurements allow us to determine the dynamic viscosity of the $\MCSRD$ fluid in homogeneous linear shear flow. Ideal velocity profiles can be obtained by periodic boundary conditions into the $x$-- and $y$--directions and either moving the $z$--walls into the $x$--direction with opposite velocities $\pm U/2$ or by applying Lees--Edwards boundary conditions at $z=\pm L/2$ in $z$--direction \cite{Lees1972}. Because undesired wall effects are present in the former method we employ Lees--Edwards boundary conditions in our simulations. Then, the dynamic viscosity $\eta$ is given by the relation
\begin{equation}
\eta = -\frac{\sigma_{zx}}{\dot{\gamma}}
\label{eq:shear}
\end{equation}
where $\sigma_{zx}$ is the tangential stress tensor component and $\dot{\gamma}=\partial_z v_{x}$ the uniform shear rate in the bulk. In our convention of the Cartesian coordinate system, the velocity field is given by $\mathbf{v}=\dot{\gamma}z\mathbf{e}_{x}$ with $\mathbf{e}_{x}$, the unit vector in $x$--direction. Local components of the stress are measured by the \oldtext{areal}\newtext{area--weighted} averaging method as described in Sec.~\ref{ssec:stresscalc}. To determine the effect of angular momentum conservation on the dynamic viscosity we study two mono--phase systems under linear shear flow. For both systems, $\MCSRDm$ and $\MCSRDp$, we consider a cubic system of size $L^3=32^3$ with average particle number $\Navg=15$ and temperature $T=5\times 10^{-3}$. Panel (b) of Fig.~\ref{fig:flow} shows the velocity profiles for $\MCSRDp$ (black) and $\MCSRDm$ (red), respectively. Because the applied shear rate $\dot{\gamma}=6.25\times 10^{-4}\:\Delta t^{-1}$ is the same for both cases, necessarily the velocity profiles $v_{x}$ are identical. Panel (c) shows the corresponding off--diagonal components of the collisional stress tensor $\boldsymbol{\sigma}^{\mathrm{(col)}}$. When angular momentum is conserved the stress tensor $\boldsymbol{\sigma}$ is symmetric. Because the kinetic contribution $\boldsymbol{\sigma}^{\mathrm{(kin)}}$ is anyway symmetric it is sufficient to examine the collisional contribution $\boldsymbol{\sigma}^{\mathrm{(col)}}$ only. As expected, the stress tensor of a $\MCSRDm$ fluid (red) is not symmetric \cite{Goetze2007,Noguchi2008,Pooley2005,Ihle2005} with $\sigma_{zx}\approx-0.5\times 10^{-3}$ and $\sigma_{xz}$ effectively zero. For the $\MCSRDp$ fluid (black) $\sigma_{xz}$ and $\sigma_{zx}$ are equal and consequently $\boldsymbol{\sigma}^{\mathrm{(col)}}$ and $\boldsymbol{\sigma}$ are symmetric. With this finding we can verify that our angular momentum conserving algorithm works correctly. Because the amplitudes of $\sigma_{zx}$ differ between $\MCSRDm$ and $\MCSRDp$ the shear viscosities derived with eqn.~\eqref{eq:shear} differ also with $\eta_{-}\approx0.85$ and $\eta_{+}\approx0.44$, respectively. This difference conforms to the idea that a larger amount of fluctuational linear momentum is available in the exchange between fluid particles in a collision cell if vorticity is not conserved. This leads to an enhanced diffusive flux of linear momentum in the sheared fluid and, thus, to a larger dynamic viscosity.

\begin{figure}
	\centering
	\includegraphics[width=0.49\columnwidth]{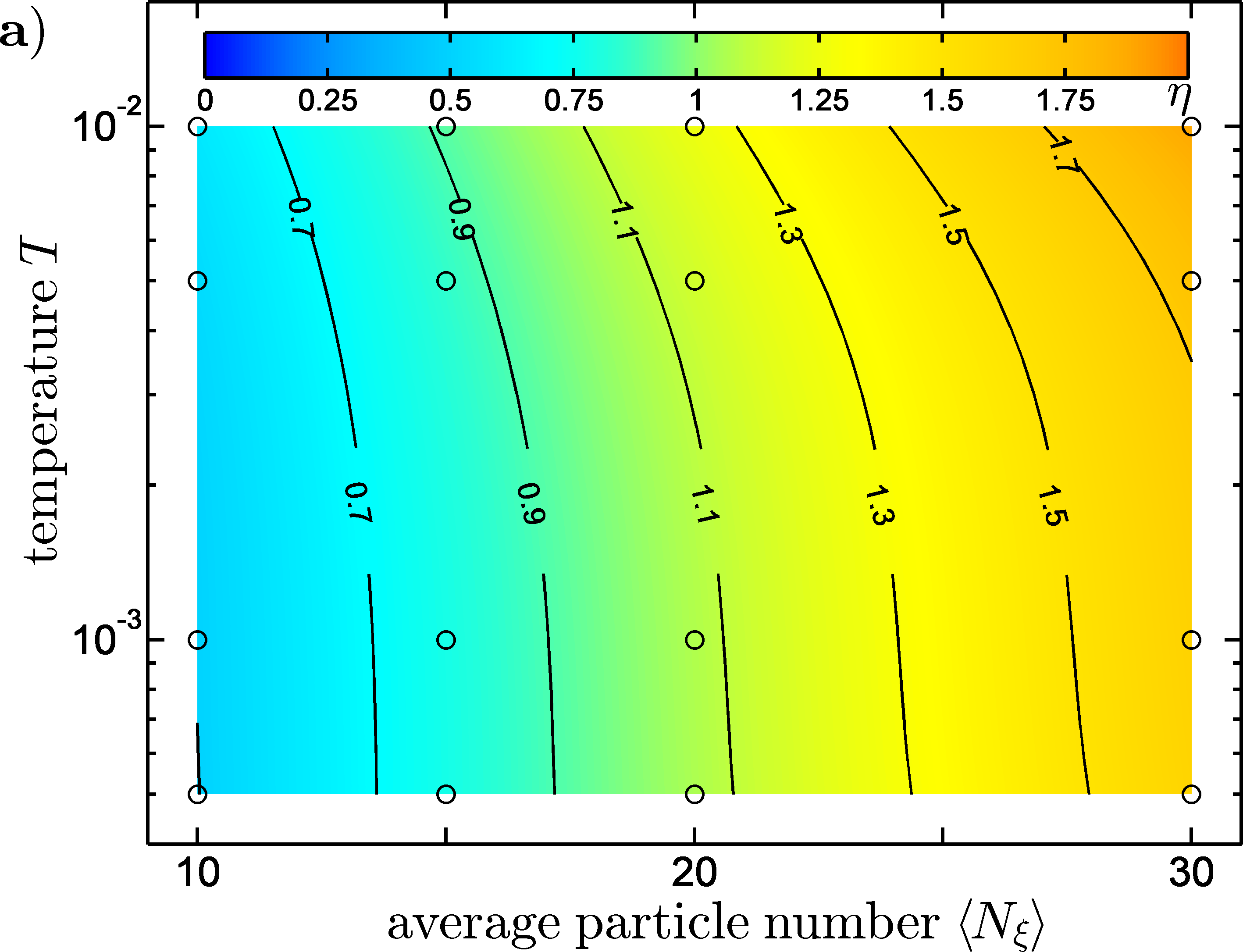}\hfill
	\includegraphics[width=0.49\columnwidth]{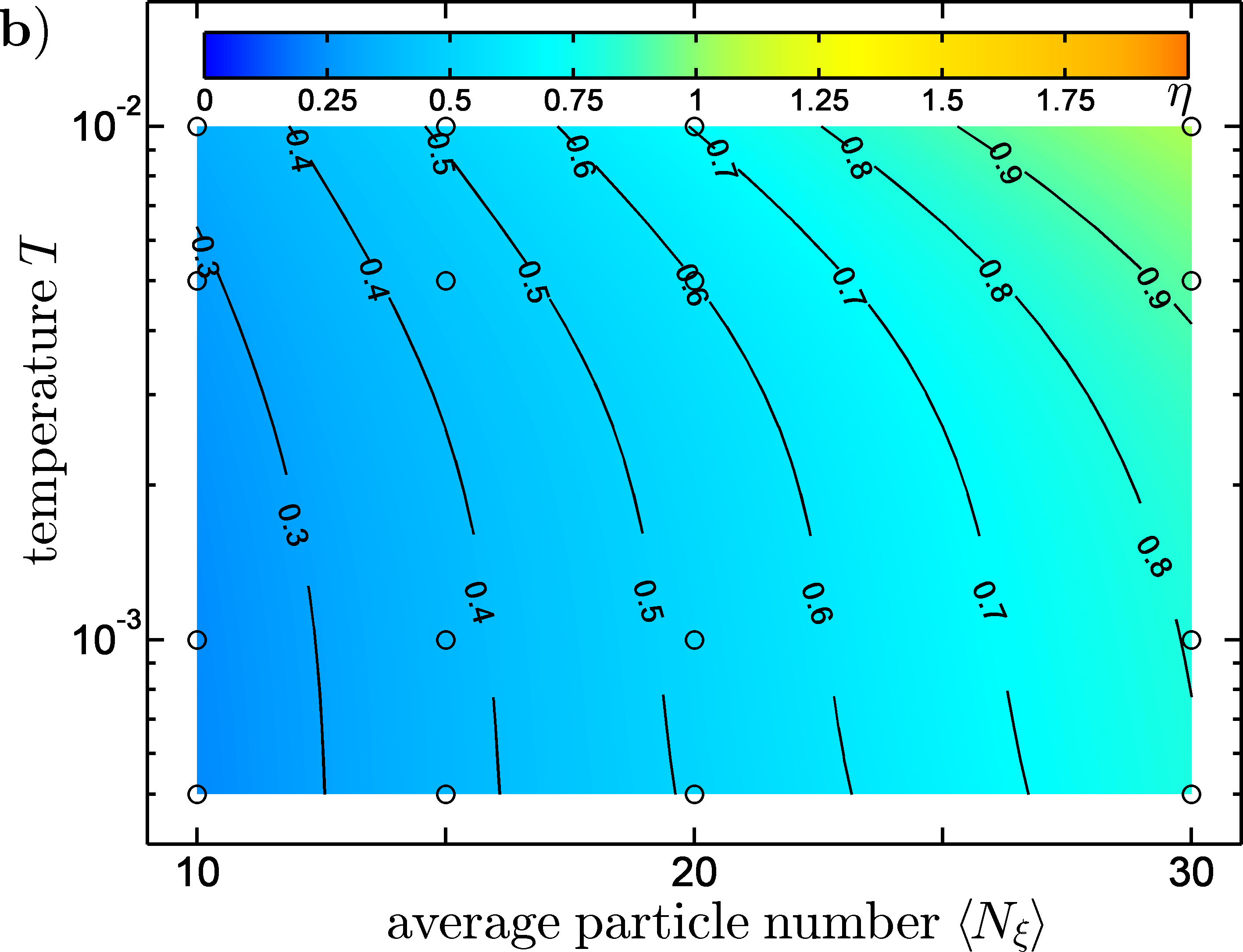}
	\caption{(color online) dynamic shear viscosity $\eta$ as a function of temperature $T$ and average particle number $\Navg$ for $\MCSRDm$ (\textbf{a}) and $\MCSRDp$ (\textbf{b}) fluids. For $\MCSRDp$ fluids the viscosity is lower by a factor of $\approx 2$.}
	\label{fig:viscmap}
\end{figure}

Figure~\ref{fig:viscmap} shows the dynamic shear viscosity $\eta_{\MCSRD}$ as a function of temperature $T$ and average particle number $\Navg$ for $\MCSRDm$ (\textbf{a}) and $\MCSRDp$ (\textbf{b}), respectively. For $\MCSRDp$ fluids all viscosities in the tested parameter range are lower by a factor of $\approx 2$. This effect of the angular momentum conservation on the viscosity of collisional dominated systems was also reported before \cite{Goetze2007}.

Although based on a similar approach, the action of a $\MCSRD$ collision operator and a standard SRD collision operator on the particle velocities are not fully equivalent. The latter collision operator employs a fixed rotation angle while the rotation angle in the former operator is determined in every collision cell from the color action principle as explained in Sec.~\ref{ssec:multiphase}. Angular momentum conservation will cause an additional departure from the dynamic viscosity of the SRD fluid given by the closed form expressions for $\eta$ eqn.~\eqref{eq:kinvisc}.

Differences between the measured dynamic viscosities of the $\MCSRDm$ or $\MCSRDp$ fluid, and the viscosity of a standard SRD fluid can be quantified using the concept of an ``equivalent'' rotation angle $\alpha_{\rm E}$ such that  $\eta_{SRD}(\alpha_{\rm E})$ according to eqn.~\eqref{eq:kinvisc} equals the measured viscosity $\eta_{\MCSRD}$. This equivalent angle depends on the average number $\Navg$ of the fluid particles and the temperature $T$. Within our tested parameter range the equivalent collision angle for $\MCSRDm$ is $\alpha_{\rm E}=(92.5\pm 2.9)^{\circ}$, whereas for $\MCSRDp$ it is lowered to $\alpha_{E}=(58.8\pm 1.4)^{\circ}$. This simply means that a mono--phase $\MCSRDm$ fluid is comparable to a standard SRD fluid with a fixed rotation angle of $\alpha\approx90^{\circ}$ and a $\MCSRDp$ fluid is comparable to a standard SRD fluid with a fixed rotation angle of $\alpha\approx60^{\circ}$, respectively. Also this finding complies perfectly to collisional dominated systems (see e.g.~\cite{Ihle2001,Ihle2003a,Gompper2008}). In this regime, the viscosity can be modified by changing the particle mass $m$ and, as we have shown, can easily be measured e.g. by a linear shear flow experiment. In a multi--phase system individual viscosities can be employed by assigning different mass $m_c$ to the fluid particles of each phase color $c$. Control over the viscosities of each single fluid phase allows us to study a wide range of problems in immiscible two--phase flows and soft condensed matter in general.

\section{Two--phase systems}
\label{sec:twophase}

Having determined the mono--phase properties of the $\MCSRD$ collision operator, we will now proceed to the substantially more complex situation of immiscible two--phase flow. Multi--phase flows in general are governed by an interplay of inertial, viscous and capillary forces. To determine the relative magnitude of these forces requires precise measurements not only of the bulk phase viscosity but also of the interfacial tension $\gamma$ between the fluid phases. A relation of the interfacial tension $\gamma$ between $\MCSRD$ fluids to the fundamental simulation parameters average particle number $\Navg$, temperature $T$, and the weights $\kappa_{cc^\prime}$ of phase colors $c,\,c^\prime$ in the collision operator is mandatory. In the following section, we will present three independent approaches to determine the interfacial tension of two coexisting phases of a $\MCSRDp$ fluid.

\subsection{Interfacial tension -- planar interface}
\label{ssec:planar}

In equilibrium the mechanical tension of a fluid--fluid interface can be expressed by an integral over the stresses in the two adjacent fluid bulk phases. Provided identical pressures in the two adjacent bulk phases, which is the case for a planar interface, we can apply the Kirkwood--Buff formula \cite{Kirkwood1949,Rowlinson1982} to calculate the interfacial tension $\gamma$ as an excess stress:
\begin{equation}
\gamma = \int \left[\sigma_{N}(x) - \sigma_{T}(x)\right]dx~.
\label{eq:KBformula}
\end{equation}
The integral in eqn.~\eqref{eq:KBformula} extends perpendicular to the interface where $\sigma_{N}(x)=\sigma_{xx}$ is the normal component and $\sigma_{T}(x)=\sigma_{yy}=\sigma_{zz}$ are the tangential components of the stress tensor $\boldsymbol{\sigma}$, respectively. The local isotropic pressure $P$ in the bulk fluid is given by $P = \Tr(\boldsymbol{\sigma})/3$ where $\Tr(\boldsymbol{\sigma})$ is the trace of the stress tensor $\boldsymbol{\sigma}$.

In mechanical equilibrium $\boldsymbol{\nabla}\cdot\boldsymbol{\sigma}=0$ holds everywhere in the fluid phases. For a planar interface between two fluids one can conclude that the component of the stress tensor normal to the interface $\sigma_{N}$ must be identical in every point of the fluid while the tangential components $\sigma_{T}$ of the stress tensor can only be functions of the coordinate $x$ normal to the interface \cite{Rowlinson1982,Walton1983,Varnik2000}. A positive interfacial tension is a requirement for spontaneous phase separation and should become visible as a depression of the tangential components $\sigma_{T}(x)$ across the interface. The normal component $\sigma_{N}(x)$, however, must be constant and identical to the respective values in the two adjacent bulk phases.

The \oldtext{areal}\newtext{area--weighted} averaging procedure of the local momentum flux according to eqns.~\eqref{eq:mopkin} and \eqref{eq:mopcol} allows the local measurement of the stress profiles across the interface of two fluids in the $\MCSRD$ model. For the sake of completeness, we will compare the results of the \oldtext{areal}\newtext{area--weighted} averaging method to those stress profiles obtained by local volume averages according to the virial approach in eqn.~\eqref{eq:virial}. The virial formulation was initially proposed for global averages but can easily be adapted to a local stress tensor measurements
\begin{equation}
\left\langle\sigma_{\alpha\beta}\right\rangle=\frac{1}{V_{\xi}}
\sum_{i=1}^{N_{\xi}}m_{i}\left\langle v_{i\alpha}{v}_{i\beta}\right\rangle-\frac{1}{V_{\xi}\Delta t}
\sum_{i=1}^{N_{\xi}}m_{i}\left\langle\left(v_{i\alpha}-v^{\prime}_{i\alpha}\right){r}_{i\beta}\right\rangle~,
\label{eq:virial2}
\end{equation}
where the volume $V_{\xi}$ is the volume of the collision cell $\xi$ and $r'$ the position of the particle relative to the center of the collision cell. In a similar manner it is possible to adjust eqn.~\eqref{eq:virial} to sample the local stress tensor on an even smaller sub--lattice, the \emph{stress grid}, to compare it with the values obtained by the \oldtext{areal}\newtext{area--weighted} averaging method.

To set up two stable $\MCSRDp$ fluid interfaces, a cubic simulation box of size $L^3=32^3$ with an average particle number of $\Navg=15$, and periodic boundary conditions are chosen. The particle species forming two immiscible phases are initialized in a planar symmetry in an ABA scheme. Without restricting generality the normal direction of the two interfaces is taken to be the $x$--direction. For symmetry reasons, the interfaces in the initial particle configuration are positioned at $x=-8a$ and $x=8a$, respectively. The temperature of the system is set to $T=5\times 10^{-3}$ so that the mean free path of a particle is then $\MFP=\Delta t \sqrt{T/m}\approx 7.1\times 10^{-2}$, and therefore the transport of linear momentum is dominated by the collisional contribution of the stress tensor. After an equilibration period of $2\times 10^{4}$ time steps the stress tensor components are averaged over $5\times 10^{4}$ subsequent time steps.

\begin{figure}
	\includegraphics[width=0.49\columnwidth]{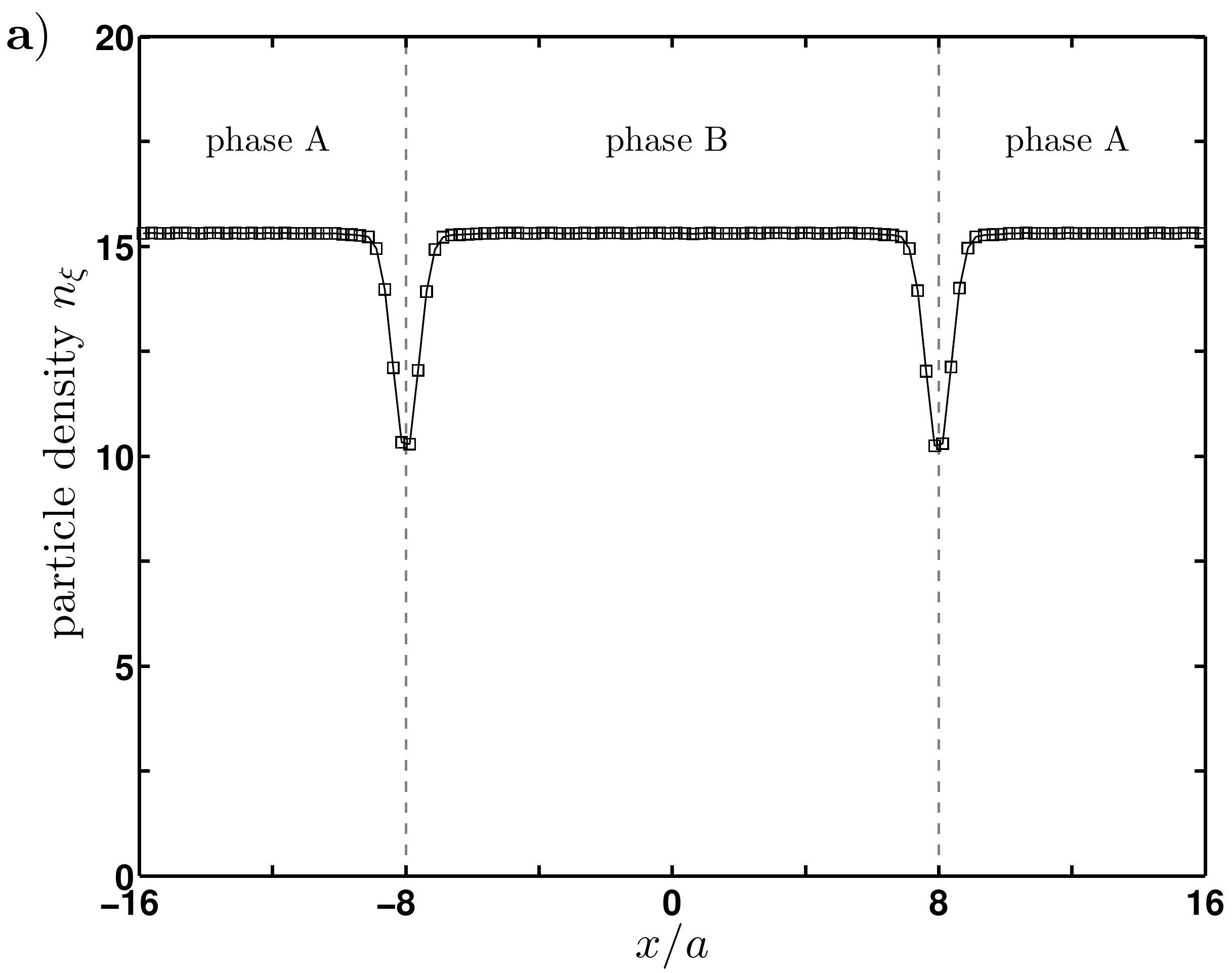}\hfill
	\includegraphics[width=0.49\columnwidth]{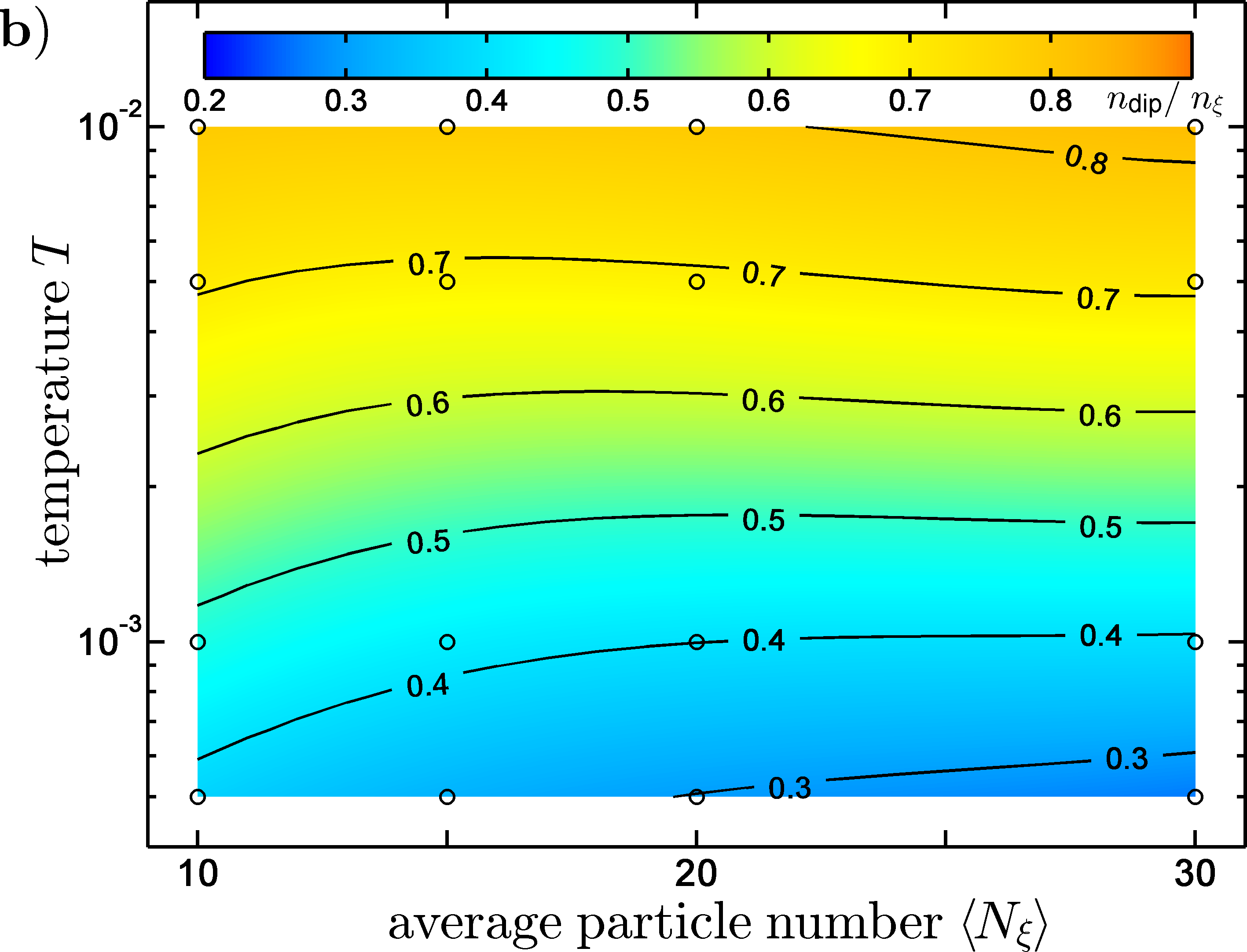}
	\caption{\small(color online) (\textbf{a}) Density profile across a system consisting of two phases A and B with planar symmetry showing a decreased particle density at the interfaces; (\textbf{b}) relative amplitude of the particle density inside the depletion layer (interface) as a function of temperature $T$ and average particle number $\Navg$.}
	\label{fig:depletion}
\end{figure}

Figure~\ref{fig:depletion}a exemplifies this ABA scheme and shows the equilibrated particle density profile in $x$--direction. Clearly visible are the two depletion zones at the interfaces (indicated by the gray dashed lines) where the local particle density is lowered to $n_{\mathrm{dip}}\approx 0.7 n_{\xi}$. Correspondingly, the particle density inside the bulk phases is increased to compensate for the decreased density in these two zones. These depletion layers originate from the $\MCSRD$ collision operator which actively drives particles of species A away from particles of species B and vice versa. Figure~\ref{fig:depletion}b summarizes the measured depression for a series of simulation runs with the same initial particle configuration but with varying average particle numbers $\Navg$ and temperatures $T$. A strong temperature dependence of the depletion is noticeable. For the lowest temperatures tested the decrease can be more than $70\%$. We will come back to these findings later in the course of this section.

\begin{figure}
	\includegraphics[width=0.49\columnwidth]{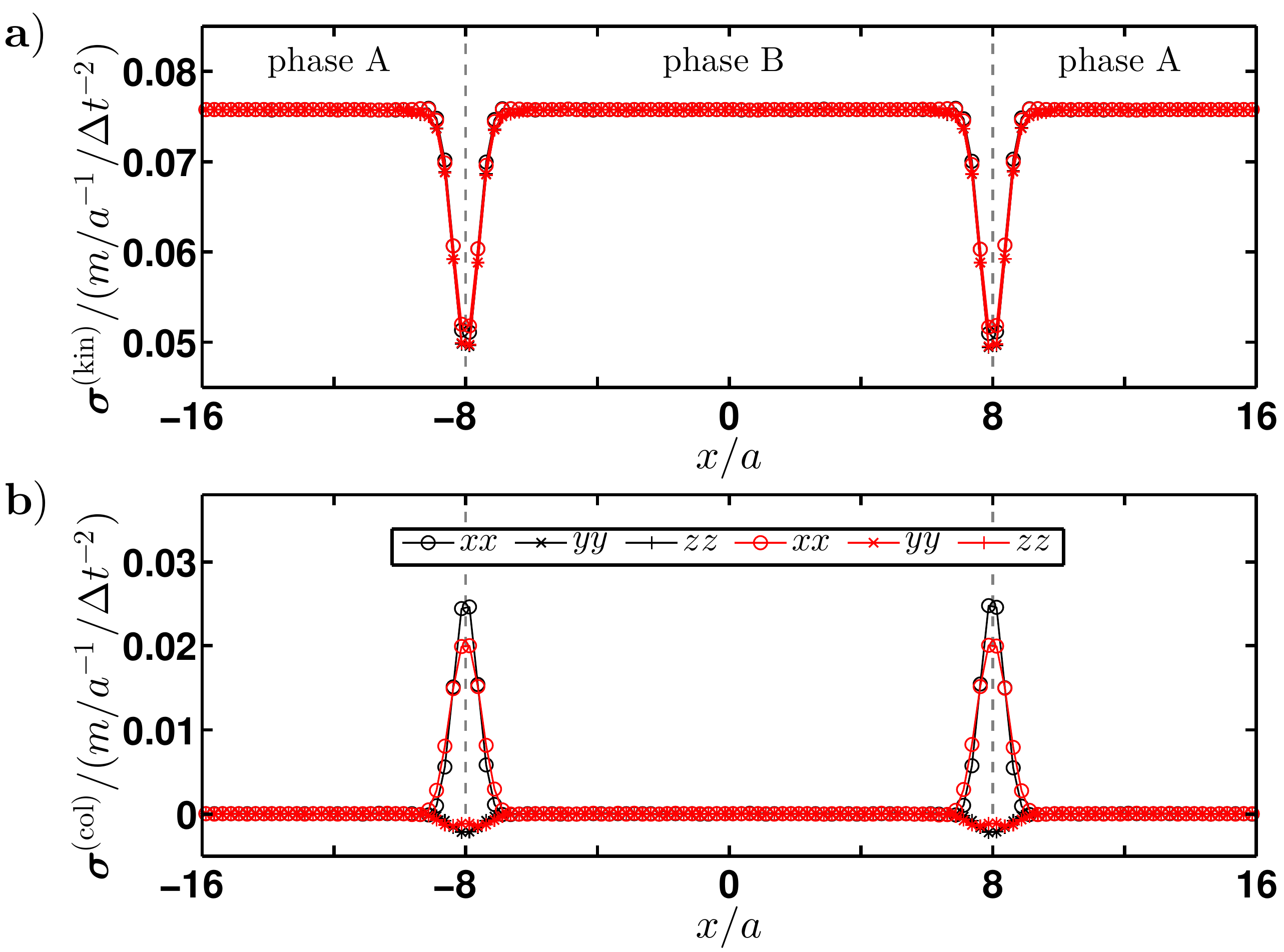}\hfill
	\includegraphics[width=0.49\columnwidth]{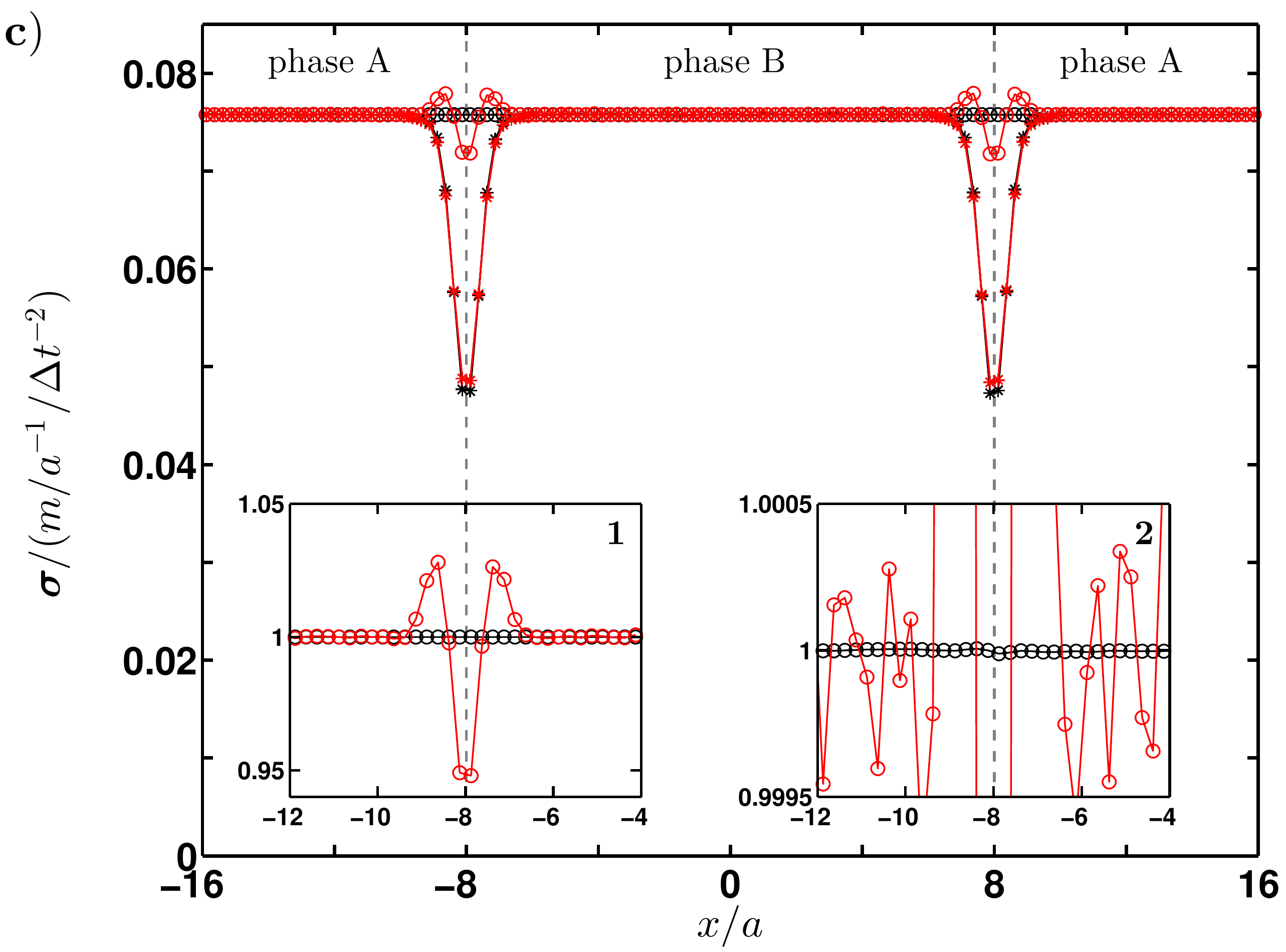}
	\caption{\small(color online) Cartesian components of the stress tensor $\sigma_{xx}$, $\sigma_{yy}$ and $\sigma_{zz}$ for a two--phase system with planar symmetry; black curves and red curves correspond to the stress measurements according to the \oldtext{areal}\newtext{area--weighted} averages eqns.~(\ref{eq:mopkin},\ref{eq:mopcol}) and volume averages eqn.~\eqref{eq:virial2}, respectively; the sub panels show the kinetic contribution (\textbf{a}), the collisional contribution (\textbf{b}) and the total stress (\textbf{c}); inset $1$ and $2$ show the normal components $\sigma_{xx}$ of both methods normalized by their bulk values, note the increased magnification for inset $2$.}
	\label{fig:stressP}  
\end{figure}

Figure~\ref{fig:stressP} shows the main diagonal components of the stress tensor obtained by \oldtext{areal}\newtext{area--weighted} averages (black) and by volume averages (red), integrated over the two translational invariant directions $y$ and $z$. Panels (a) to (c) of Fig.~\ref{fig:stressP} show the kinetic contribution, the collisional contribution, and the total stress, respectively. The distance between the sampling points is $d_{\sigma}=a/4$ which implies that the stress components in every single collision cell are sampled at $64$ individual points (see Sec.~\ref{ssec:stresscalc}).

Following the argument that the normal component of the stress tensor $\sigma_{N}(x)$ must be constant across both interfaces, it is expected that in equilibrium the flux of particles away from the interface driven by the multi--color collision operator equals the diffusive flux of bulk particles towards the interface. Otherwise the position and density profile of the interface could not be stationary. For both methods the kinetic contribution of all three main components has a dip across the interface and the values from the \oldtext{areal}\newtext{area--weighted} and volume averaging methods are almost identical (Fig.~\ref{fig:stressP}a). In contrast, the collisional contributions for both methods shown Fig.~\ref{fig:stressP}b are zero inside the bulk. Only the normal components $\sigma_{xx}(x)$ display a peak at the interface positions $x=-8a\:,x=8a$. The tangential components $\sigma_{yy}(x)$ and $\sigma_{zz}(x)$, however, display a small dip at the interface. Clearly visible is the difference between the peaks for the normal components $\sigma_{xx}(x)$ of the \oldtext{areal}\newtext{area--weighted} averaging (black circles) and of the volume averaging methods (red circles).

The total interfacial stress being the sum of both contributions is plotted in Fig.~\ref{fig:stressP}c. As required for a stable interface in mechanical equilibrium, the tangential components $\sigma_{yy}$ and $\sigma_{zz}$ exhibit a dip across the interfaces. The dip in the example shown in Fig.~\ref{fig:stressP}c is caused mainly by a dip in the kinetic contribution of the stress tensor. In contrast to the \oldtext{areal}\newtext{area--weighted} averaging method where the normal component $\sigma_{xx}$ (black circles) is constant across the interface, the volume averaging method shows a \emph{non--physical fluctuation} in $\sigma_{xx}$ (red circles). Inset $1$ of Fig.~\ref{fig:stressP}c shows the normal components for both methods normalized by their average value inside the bulk. The over-- and undershoot of the volume averaging method covers a range of $\approx\pm5\%$. Even for a magnification which is higher by two orders of magnitude as shown in the inset $2$ of Fig.~\ref{fig:stressP}c there is no such effect visible for the \oldtext{areal}\newtext{area--weighted} averaging method so that this description complies significantly better to the expected constancy of $\sigma_{xx}(x)$.

This \emph{unphysical} behavior of the stress profile obtained by the volume averages according to eqn.~\eqref{eq:virial2} can be explained from the particular choice of the point where the momentum exchange is localized during a multi--particle collision. A consistent measurement of the stress tensor components is not possible within the volume averaging approach. Despite this obvious drawback, we like to point out that the values for the surface tension for both methods differ by less than $0.5\%$. In $\MCSRD$ simulations where only an integral value of the isotropic pressure $P$ is required it is computationally advantageous to employ the volume averaging approach. Another rather technical aspect is the choice of the suitable resolution of the \emph{stress grid}. For the purpose of the presented simulations, a factor of $4$ gave an optimal trade--off between the spatial resolution and the additional computational overhead that is required for the \oldtext{areal}\newtext{area--weighted} averaging method.

\begin{figure}
	\includegraphics[width=0.49\columnwidth]{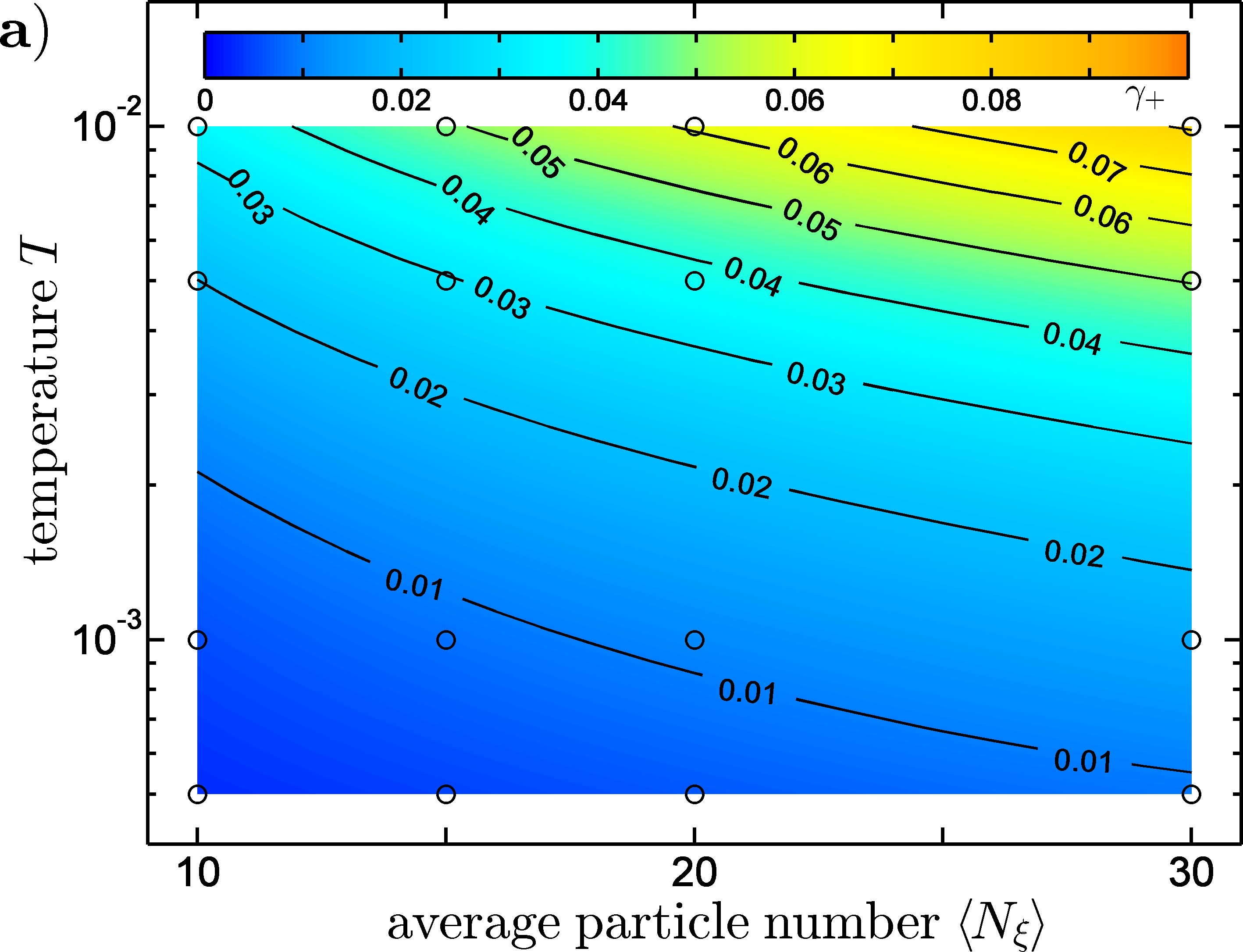}\hfill
	\includegraphics[width=0.49\columnwidth]{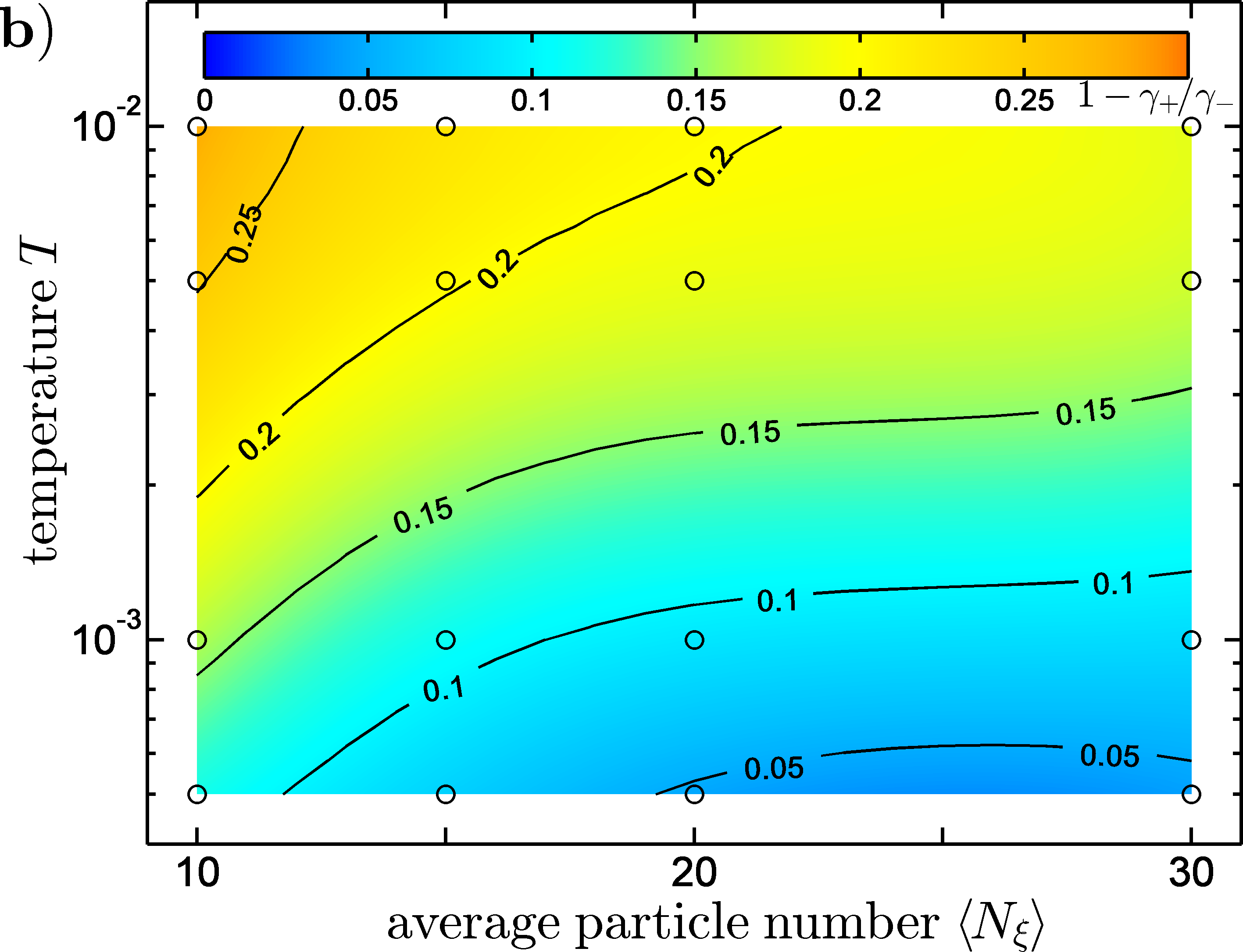}
	\caption{(color online) panel (\textbf{a}) shows the interfacial tension $\gamma$ as a function of temperature $T$ and average particle number $\Navg$ for $\MCSRDp$ fluids; in panel (\textbf{b}) the ratio $1-\gamma_{+}/\gamma_{-}$ between the interfacial tension for $\MCSRDp$ and $\MCSRDm$ fluids is shown.}
	\label{fig:tensmap}
\end{figure}

Figure~\ref{fig:tensmap} shows the interfacial tension derived with the Kirkwood--Buff formula eqn.~\eqref{eq:KBformula} over a range of control parameters $\Navg$ and $T$. While panel (a) shows the interfacial tension for a $\MCSRDp$ fluid, panel (b) shows the relative difference of the interfacial tension between $\MCSRDp$ and $\MCSRDm$ fluids, respectively.
For the considered control parameters the interfacial tension for $\MCSRD$ fluids increases with increasing $\Navg$ and $T$ where the dependence on $T$ is stronger than on $\Navg$. The values for $\MCSRDm$ fluids differ by $\approx 25\%$ compared to those of $\MCSRDp$ fluids if $T$ is high and $\Navg$ is low. The lower the temperature and the higher the average particle number, the smaller is the difference between the interfacial tension of $\MCSRDp$ and $\MCSRDm$ fluids ($<5\%$).
 
The increase of interfacial tension with increasing temperature correlates well with the \emph{strength} of the depletion layer created at an interface (see Fig.~\ref{fig:depletion}). If the temperature is low and correspondingly the mean free path of the particles is short a smaller interfacial tension is needed to counterbalance the flux of particles towards the interface. Similarly, if the temperature is high a stronger interfacial tension is needed to counterbalance the particle flux. This also means that it is not possible to arbitrarily tune the interfacial tension in a $\MCSRD$ system. If the mean free path is too short (low temperatures in our case) it may happen that collision cells contain no particles and therefore any hydrodynamical behavior is lost. On the contrary, if the mean free path is too long (high temperatures in our case) the phase segregation is no longer stable. Particles of one phase may end up after the velocity update and next streaming step inside the other phase and effectively rupture the interface. If we assume that the interface is stable up to a mean free path of $\MFP\approx0.5a$ then the upper limit for the system temperature in our setup would be $T\approx0.25$. The effect of the depletion layer is addressed again in Sec.~\ref{ssec:twophasedriving} when we study the slip between two fluid phases. We like to point out that the phase segregation of the $\MCSRD$ operator is most effective if thermal fluctuations are low and the system is in the collisional--dominated regime and therefore the collisional part in the stress tensor dominates over the advective part.

\subsection{Interfacial tension -- Young--Laplace equation}
\label{ssec:laplace}

An alternative to the interfacial tension measurements from a microscopic stress profile employs the Young--Laplace equation $\Delta P=2\gamma/R$, where $R$ is the radius of a spherical droplet of fluid A in mechanical equilibrium with the ambient fluid B. Measurements of the difference $\Delta P \equiv P_A-P_B$ of bulk pressures $P_i$ in the fluids $i=A,B$ allows us to infer the interfacial tension $\gamma$ from the constant of proportionality between $\Delta P$ and the curvature $R^{-1}$ of the interface. This method has been used before to determine $\gamma$ in two--phase SRD fluids \cite{Inoue2004,Tuezel2007}

\begin{figure}
	\centering
	\includegraphics[width=0.49\columnwidth]{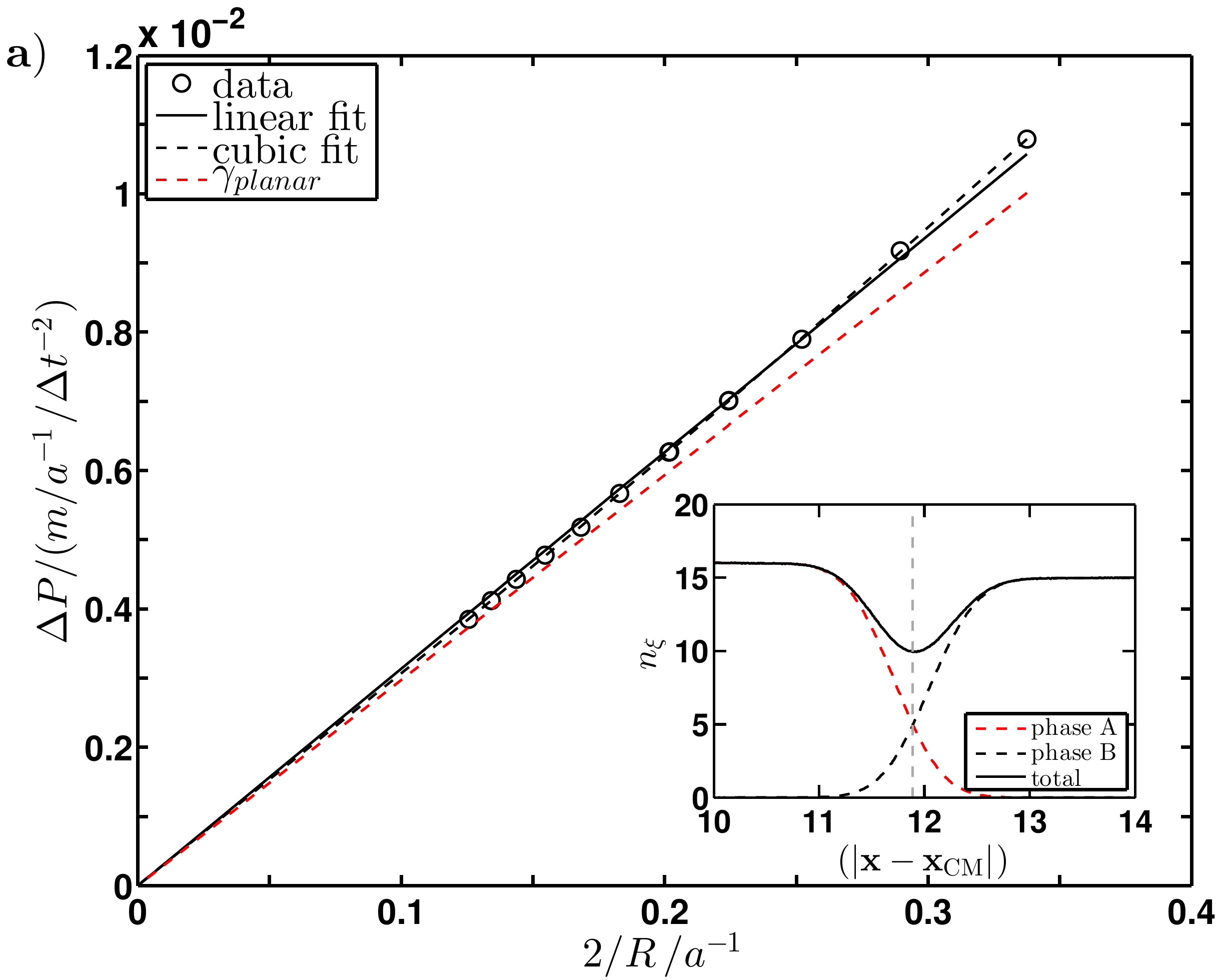}\hfill
	\includegraphics[width=0.49\columnwidth]{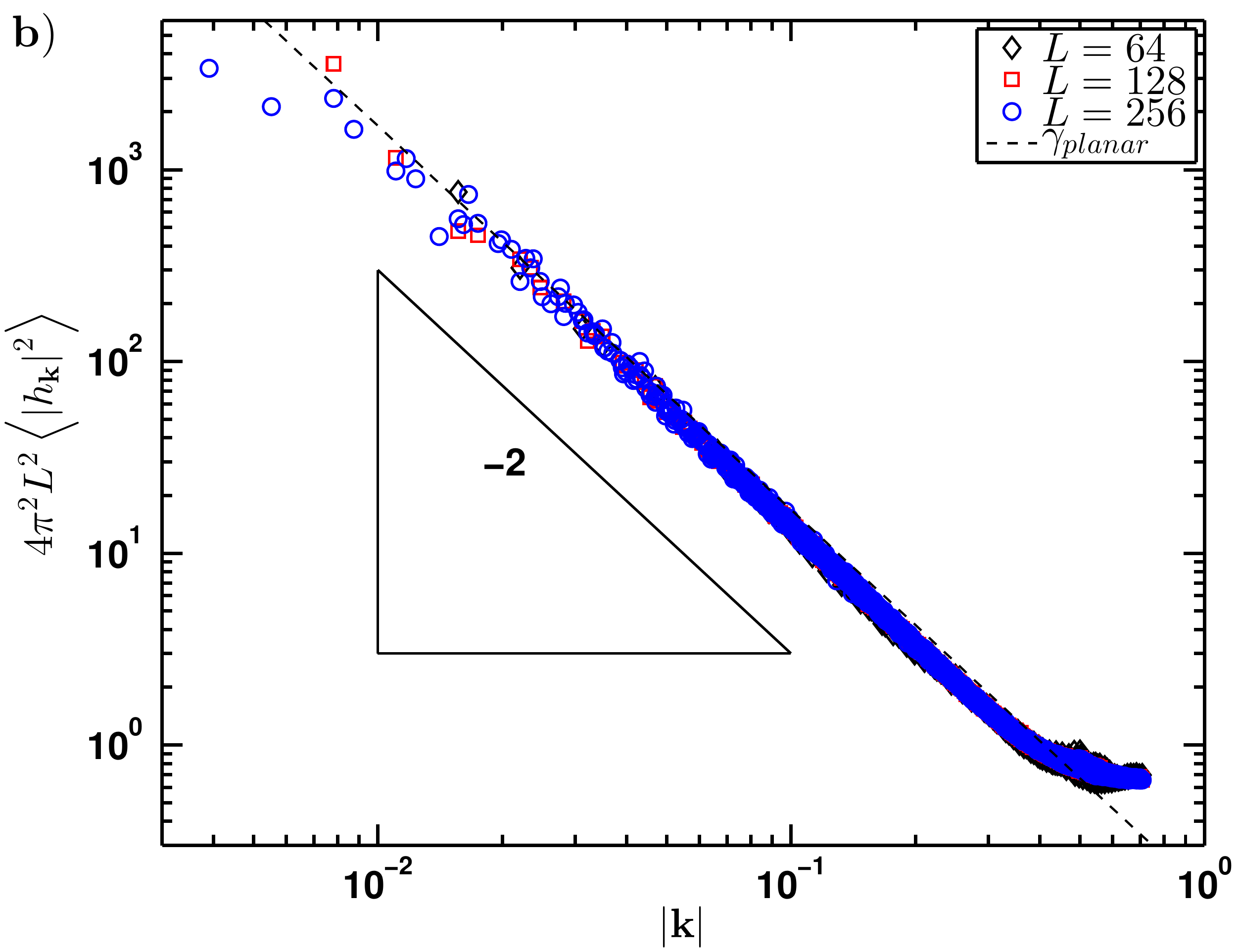}
	\caption{(color online) (\textbf{a}) Pressure difference $\Delta P$ as a function of drop radius $R$ (the error bars are smaller than the symbol size), the inset shows the particle density $n_\xi$ as a function of the radial distance to the drop center of mass; \textbf{b} Capillary wave spectrum for a three systems with different sizes $L$ employing the same temperature $T$ and average particle number $\Navg$ as the system used in (a).}
	\label{fig:laplace}
\end{figure}

We perform the Young--Laplace test as a further benchmark of the values obtained by the planar interface method outlined in Fig.~\ref{fig:tensmap}. Due to high symmetry of the droplet, we employ a cubic simulation box of size $L^3=64^3$ with periodic boundary conditions, and apply the standard control parameter temperature $T=5\times 10^{-3}$ and average particle number $\Navg=15$. Different sized spherical droplets of fluid A are placed in the center of the box while the remaining space is uniformly filled with particles of fluid B. In line with the simulations described before (Sec.~\ref{ssec:planar}), we obtain time averages of all measured quantities for a duration of $5\times 10^{4}$ time steps. Before this measurement interval we waited for $2\times 10^{4}$ time steps to ensure a sufficient equilibration of the droplet and the ambient fluid. During the measurement interval the pressure $P_A$ and $P_B$ in the bulk fluids is determined by the \oldtext{areal}\newtext{area--weighted} averaging method as described in Sec.~\ref{ssec:stresscalc}. The results are shown in Fig.~\ref{fig:laplace}a.

Due to the Brownian motion of the droplet, we displace all fluid particles in the simulation box in regular intervals by a shift $\mathbf{d}_{c}=\mathbf{x}_{\mathrm{CM}}-\mathbf{x}_{\mathrm{CS}}$ where $\mathbf{x}_{\mathrm{CM}}$ is the center of mass position of all fluid particles in the droplet, and $\mathbf{x}_{\mathrm{CS}}$ the center of the simulation box. This procedure avoids `smearing out' of the relevant physical quantities by a diffusion of the droplet's center of mass. The inset in Fig.~\ref{fig:laplace}a shows the particle density as a function of the radial distance from the center of mass $\left|\mathbf{x}-\mathbf{x}_{\mathrm{CM}}\right|$. The dashed lines are the individual color densities and the solid black line is the total particle density, respectively. The radius of the equilibrated droplet is taken to be the crossing point of the individual color densities (vertical gray dashed line). The width of the interface is approximately one lattice unit which is the intrinsic length scale determined by the collision operator. The inset of Fig.~\ref{fig:laplace}a also illustrates the density difference between the droplet and the ambient bulk phase created by the self--compression of the droplet phase.

In the example above the initial radius of the droplet is $R_{\mathrm{i}}=12a$. Because of self--compression the particle density inside the droplet is approx. $7\%$ higher and the final radius of the equilibrated droplet is $R_f=11.89a<R_i$. Due to the fact that a $\MCSRD$ fluid is rather a gas than a liquid the smaller the droplet radius and therewith the larger the curvature of the droplet, the larger is the density difference between the droplet and the bulk. The density difference between the two phases is directly linked to the pressure difference by the ideal gas law. We will account for the density difference in the course of this section when we determine the interfacial tension from the Young--Laplace equation. In Fig.~\ref{fig:laplace}a the pressure difference $\Delta P$ between droplet and bulk is shown as a function of the inverse equilibrated radius $R^{-1}$. The solid black line displays the result of a linear fit to the simulation data. The interfacial tension derived from this fit deviates by almost $10\%$ from the interfacial tension measured with the planar interface method (red dashed line).

To account for the dependence of the interfacial tension $\gamma$ on the particle densities $n_A$ and $n_B$ in the adjacent bulk fluids, we expand $\gamma(n_A,n_B)$ in a Taylor series in powers of the density difference $\Delta n=n_A-n_B$:
\begin{equation}
\gamma(n_A,n_B) = \gamma_0(\bar n) + C(\bar n)\,\Delta n^2+{\cal O}\left(\Delta n^4\right),
\label{eq:laplace1}
\end{equation}
where $\bar n=(n_A+n_B)/2$ is the average particle number. We can readily identify $\gamma_{0}(\bar n)$ as the interfacial tension of a plane interface between two bulk phases of equal density $\bar n$ and an empirical constant $C(\bar n)$. Both functions $\gamma_0(\bar n)$ and $C(\bar n)$ have to be determined from the simulation data. Odd terms in the Taylor expansion in eqn.~\eqref{eq:laplace1} must vanish because the symmetry $\kappa_{AB}=\kappa_{BA}$ of weights in the $\MCSRD$ collision operator implies $\gamma(n_A,n_B)=\gamma(n_B,n_A)$.

Using the relation $n_i= P_i/k_BT$ for an ideal gas, we can	rewrite the Young--Laplace equation as an implicit equation
\begin{equation}
k_B T \Delta n = \frac{2\gamma\left(n_A,n_B\right)}{R}
\end{equation}
in the density difference $\Delta n$ which can be solved with expansion eqn.~\eqref{eq:laplace1} in form of a power series in the curvature $R^{-1}$. After reexpressing the density difference by the pressure difference, we finally arrive at the relation
\begin{equation}
\Delta P = \frac{2\gamma_{0}}{R}\left(1 + \frac{4C\gamma_0}{(k_{B}T)^2\,R^{2}}\right)+\mathcal{O}\left(R^{-5}\right)~.
\label{eq:laplace4}
\end{equation}
In the present study, we neglect all higher order terms $\sim{\cal O}(R^5)$ and fit the cubic expression eqn.~\eqref{eq:laplace4} in the curvature $R^{-1}$ to the simulation data, cf.~the dashed black line in Fig.~\ref{fig:laplace}a. The zeroth order value $\gamma_{0}$ deviates by less than $0.4\%$ from the interfacial tension $\gamma_{planar}$ measured from the microscopic stress profile of the planar interface. As expected, for large droplets and hence small curvatures, the interfacial tension values approach the value for the planar interface in the asymptotic limit of zero curvature. Only the relation in eqn.~\eqref{eq:laplace4} accounts for the increased density difference when considering small droplets which is reflected in the quality of the fit.

We determined the zeroth order interfacial tension $\gamma_0$ from fits according to eqn.~\eqref{eq:laplace4} for the range of control parameters $T$ and $\Navg$ used in the computation of $\gamma_{planar}$ shown in Fig.~\ref{fig:tensmap}a. The maximum relative difference between $\gamma_{planar}$ and $\gamma_0$ with $(0.8\pm0.4)\%$ turns out the be very small.

\subsection{Interfacial tension -- thermal fluctuations}
\label{ssec:thermal}

Interfacial tension counteracts an increase of surface area and, thus, suppresses the roughening of a fluid--fluid interface by thermal fluctuations \cite{Safran1994,Flekkoy1995,Flekkoy1996}. Fourier amplitudes $h_\mathbf{k}$ of the displacement field $h(\mathbf{r})$ measured with respect to an initially flat configuration contribute according to the equipartition theorem
\begin{equation}
	\left\langle \left| h_{\mathbf{k}} \right|^{2} \right\rangle = \frac{k_{B}T}{4\pi^2 L^2 \gamma \left| \mathbf{k} \right|^{2}},
\label{eq:fourier}
\end{equation}
leading to an RMS roughness proportional to $\sqrt{k_{B}T/\gamma}$. Figure~\ref{fig:laplace}b displays averages $\langle |h_\mathbf{k}|^{2}\rangle$ of an expansion into planar capillary waves as a function of the wave number $k$. The system parameters for this example are again $T=5\times 10^{-3}$ and average particle number is $\Navg=15$. Apart from the cut off at large wave vectors $k_{max}\approx \pi/a$, the data points for the three different simulation box sizes $L^3 \in \{64^3,128^3,256^3\}$ conform to the characteristic power--law scaling in eqn.~\eqref{eq:fourier} in $k$ with an exponent $-2$. The dashed line shows the expected relation eqn.~\eqref{eq:fourier} with the value $\gamma_{planar}$ from the interfacial tension measurement. Deviations from the ideal scaling for small wave numbers can be attributed to the poor statistics for large wave lengths.

We used three independent methods to determine the interfacial tension between two $\MCSRD$ fluids. We could show that the model correctly reproduces thermal fluctuations of the interface and that the interfacial tension can readily be determined with one of the above methods for any given set in the range of control parameters tested.

\subsection{Interfacial slip}
\label{ssec:twophasedriving}

\begin{figure}
	\centering
	\includegraphics[width=0.49\columnwidth]{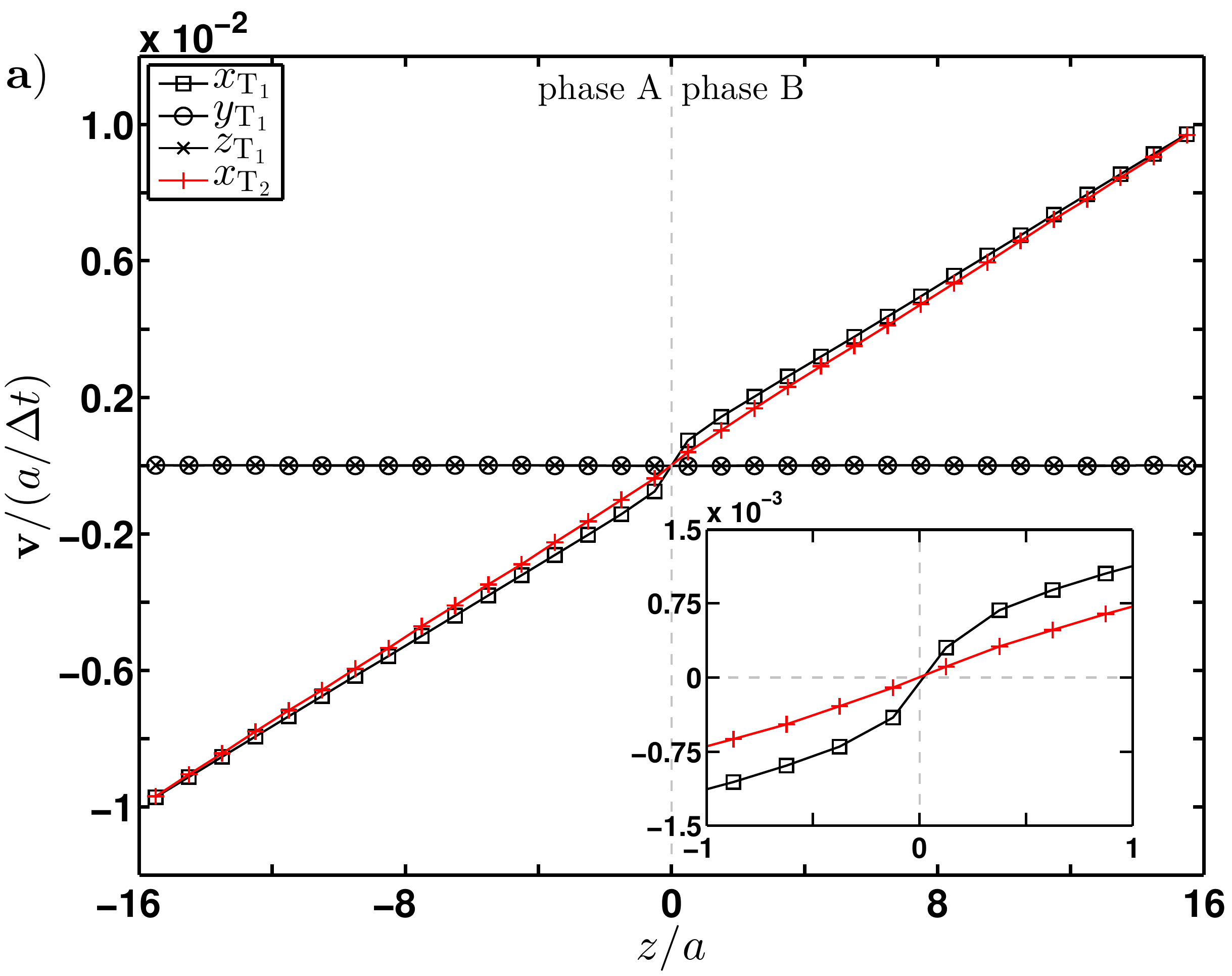}\hfill
	\includegraphics[width=0.48\columnwidth]{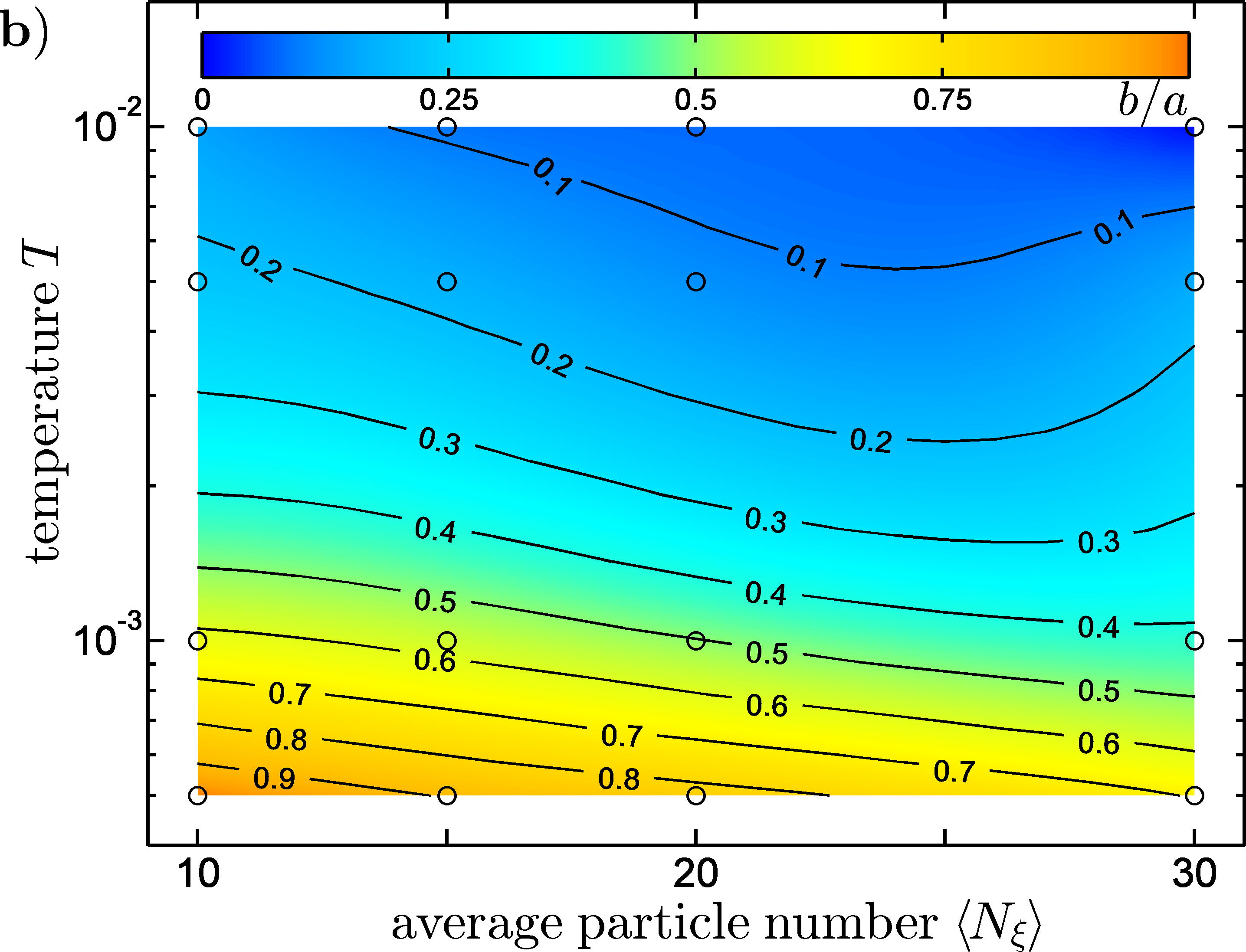}
	\caption{(color online) \textbf{a)} shear profile velocity components of a two--phase system for two different temperatures $\mathrm{T_1}=5\times 10^{-4}$ (black) and $\mathrm{T_2}=5\times 10^{-3}$ (red), the spatial sampling rate inside the inset is higher by a factor of 4; \textbf{b)} Interfacial slip $b$ as a function of system temperature $T$ and average particle number $\Navg$.}
	\label{fig:slip}
\end{figure}

As we have shown before for the planar symmetry (Sec.~\ref{ssec:planar}) and the Young--Laplace test (Sec.~\ref{ssec:laplace}), a depletion layer of lower particle density develops between two $\MCSRD$ phases. This is due to the very nature of the multi--color algorithm and cannot be avoided. To elucidate how this depletion layer effects the interface behavior we perform again a shear experiment as described in Sec.~\ref{ssec:viscosity}. In contrast to the mono--phase experiments this time the system is half--filled with phase A ($z_{A}\in[-L_{z}/2,0]$) and half--filled with phase B ($z_{B}\in{]0,L_{z}/2]}$). Lees--Edwards boundary conditions are again applied in $z$--direction. The parameter range is the same as used before and angular momentum conservation is switched on ($\MCSRDp$).

Figure~\ref{fig:slip}a shows velocity profiles of two examples out of the set with two different temperatures ($\mathrm{T_1}=5\times 10^{-4}$ and $\mathrm{T_2}=5\times 10^{-3}$). For the sake of clarity, only the $y$-- and $z$--components of the velocity are indicated for $\mathrm{T_1}$. The inset in Fig.~\ref{fig:slip}a magnifies the $x$--component of the velocity profiles around $z=0$. Clearly visible is the offset for $\mathrm{T_1}$ (black curve) corresponding to an interfacial slip between the two fluid phases. An extrapolation length similar to the Navier slip length can be defined by first extrapolating the velocity profiles in both bulk fluids to the lateral position $x_0^A$ and $x_0^B$ with $v_{z}(x_0^{A,B})=0$, and then defining the slip length $b$ as the difference $b=x_0^A-x_0^Ba$. For a small temperature $\mathrm{T_1}$, the interfacial slip length is $b=(0.89\pm0.02)a$ which is in the order of the spatial range of the $\MCSRD$ operator. When increasing the temperature by an order of magnitude to $\mathrm{T_2}$ (red curve) the interfacial slip length reduces significantly to $b=(0.17\pm0.01)a$. This weak slip is almost no longer visible in Fig.~\ref{fig:slip}a as well as in the inset. Figure~\ref{fig:slip}b shows the interfacial slip as a function of average particle number $\Navg$ and temperature $T$ for the complete parameter range. As expected from the planar symmetry test (Sec.~\ref{ssec:planar}) the interfacial slip depends much stronger on temperature than on average particle number. At high temperatures the amplitude of the depletion layer is rather small and momentum can effectively be transported across the interface leading to a reduced slip length $b$. Therefore, the same arguments hold for the interfacial slip as well as for the slip on solid walls. An increase in temperature leads to a decrease of the slip length and vice versa \cite{Bolintineanu2012}.

A good trade--off between the unphysical slip at the interface and a possible rupturing is reached at temperatures of $T\approx5\times 10 ^{-3}$. Hence, we will use this temperature value throughout the following simulations.

\subsection{Drop in linear shear flow}
\label{ssec:taylor}

Multi--phase flows are characterized by an interplay of inertial, viscous, and capillary forces. The relative magnitude of these forces is described by two non--dimensional numbers, the capillary number $Ca\equiv\eta U/\gamma$ and the Reynolds number $Re\equiv \rho U L^{\ast}/\eta$ where $U$ and $L^{\ast}$ are the characteristic velocity and length scale of the flow, respectively. In the regime of small $Ca\ll1$, capillary forces dominate over viscous forces while small Reynolds numbers $Re\ll 1$ indicate that inertia can be neglected as compared to viscous forces. In the limit $Re \ll 1$, we can assume that capillary and viscous stresses are in equilibrium at any point in time.

To test whether the $\MCSRDp$ algorithm correctly reproduces the interplay between capillary, viscous and inertial forces, we studied the steady deformation of a viscous drop subjected to a linear shear flow. The deformation of drops in different flow fields was first studied in the pioneering experiments of Taylor~\cite{Taylor1932,Taylor1934}. Depending on the viscosity ratio, the drop not only deforms under the influence of the incident shear flow but also disintegrates above a certain shear rate into smaller daughter droplets. The dynamics of drop breakup in shear flow was studied experimentally and numerically by e.g.~\cite{Bentley1986,Rallison1981,Rallison1984}. It is beyond the scope of this work to examine the whole variety of deformation and breakup patterns in view of the particular predictions of the models but we refer the reader to the overview work of Ref.~\cite{Stone1994}.

\begin{figure}
	\centering
	\includegraphics[width=0.8\columnwidth]{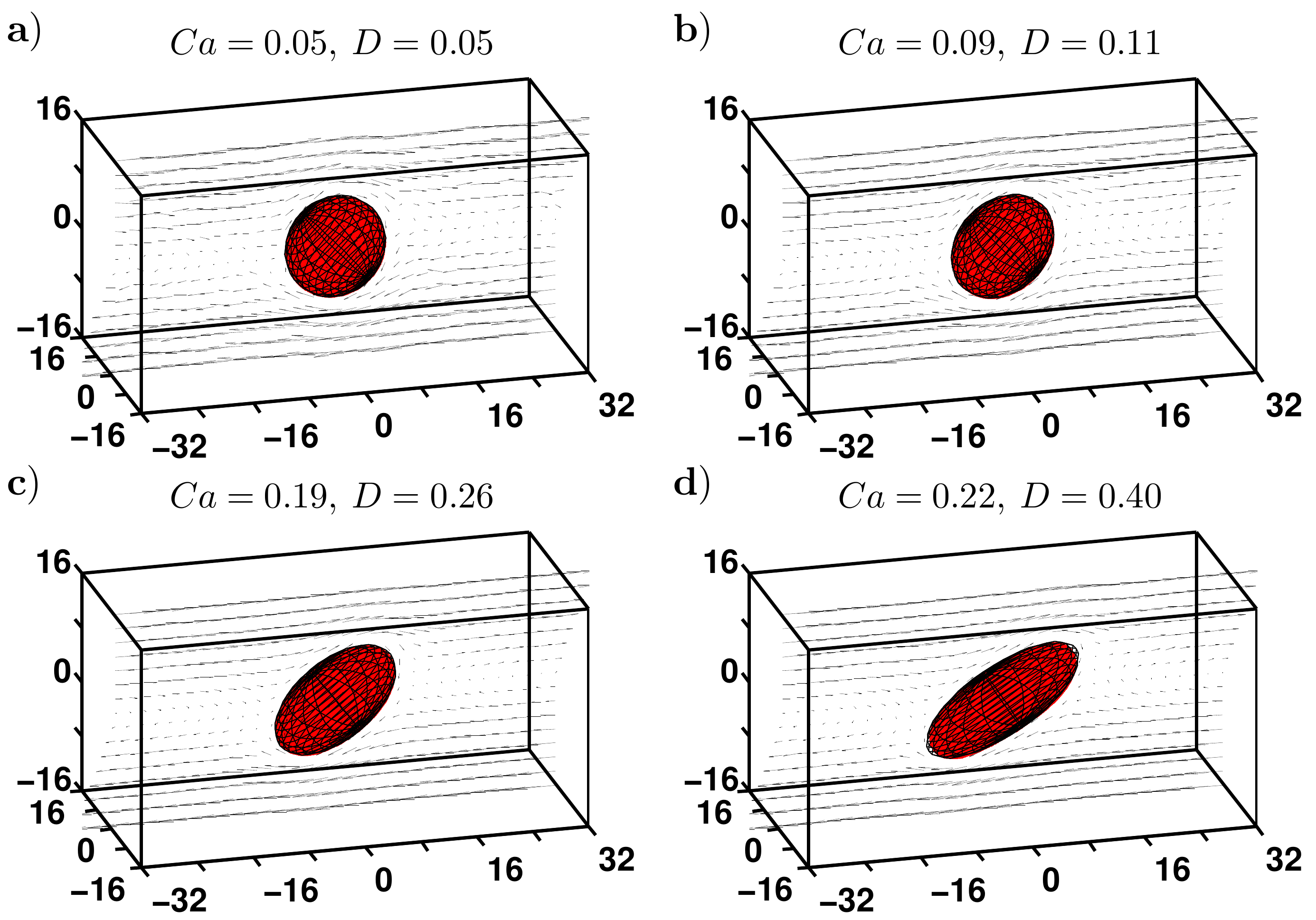}
	\includegraphics[width=0.8\columnwidth]{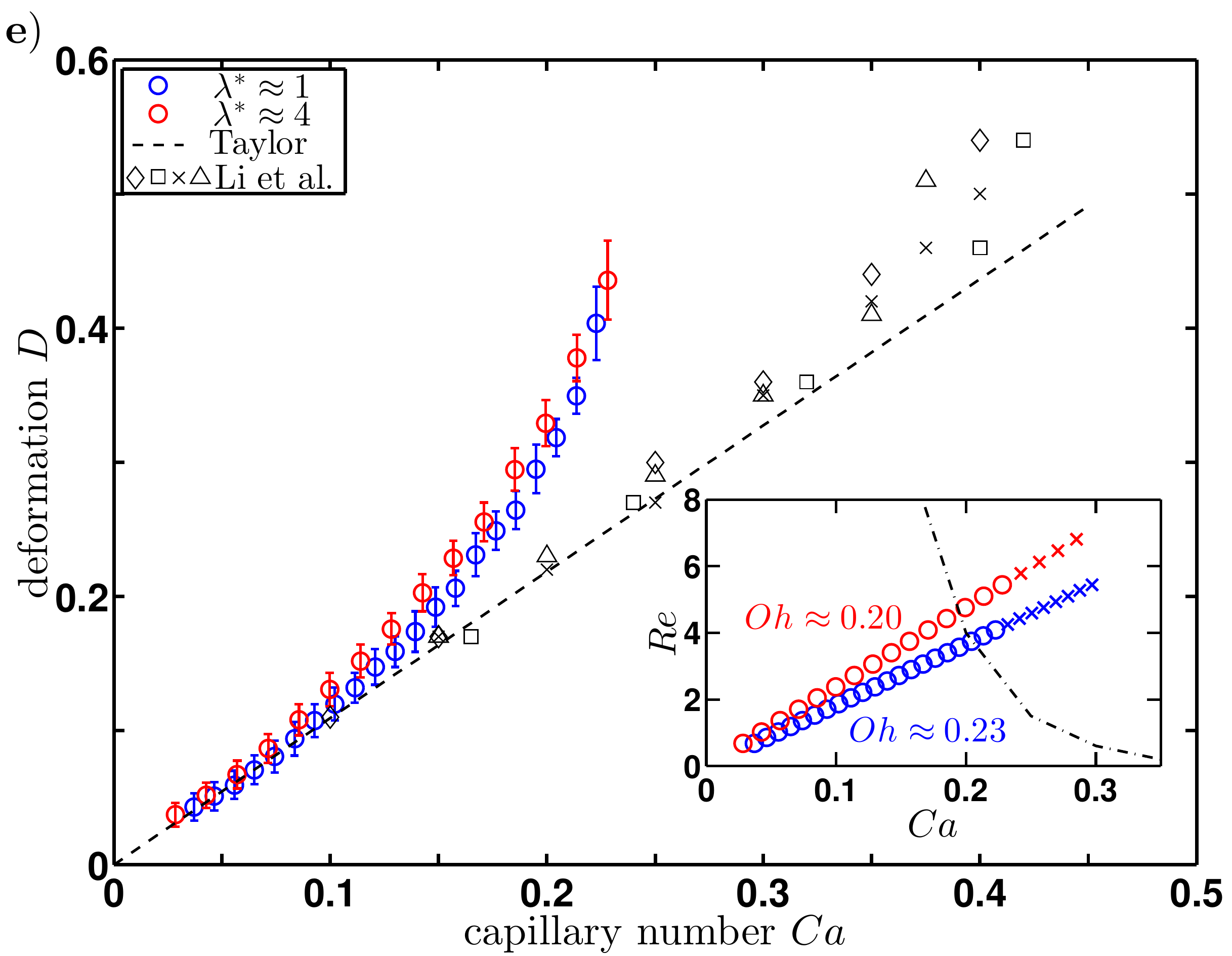}
	\caption{(color online) (\textbf{a-d}) steady state drop shapes in linear shear flow for four different capillary numbers $Ca$; the time averaged interface is shown in red and the corresponding ellipsoidal fit as grid overlay; gray arrows indicate the velocity field in the center of the domain (\textbf{e}) deformation $D$ as function of capillary number $Ca$ for viscosity ratios $\lambda^{\ast}\approx 1$ (blue) and $\lambda^{\ast}\approx 4$ (red); \newtext{the black dashed line and symbols are taken from Ref.~\cite{Li2000} and refer to deformations in the Stokes flow limit of $Re=0$}; the inset shows all simulations in the $Re$--$Ca$ phase plane, circles and crosses indicate stable and unstable droplet configurations, respectively.}
	\label{fig:taylor}
\end{figure}

Deformations of a single droplet in a linear shear flow depend on the radius $R$ of the undeformed, spherical drop and its dynamic viscosity $\eta_{d}$ and mass density $\rho_{d}=m_d\,n$, the density $\rho_{b}=m_b\,n$ and the dynamic viscosity $\eta_{b}$ of the ambient fluid phase, the interfacial tension $\gamma$, and the asymptotically reached shear rate $\dot{\gamma}=\partial_z v_{x}$ far away from the droplet. Hence, the relevant dimensionless control parameters are the ratio of drop and bulk viscosities $\lambda^{\ast}\equiv\eta_{d}/\eta_{b}$, the capillary number $Ca\equiv \dot{\gamma}\eta_{b} R/\gamma$ and the Reynolds number $Re=\dot{\gamma}\rho_{b} R^{2}/\eta_{b}$. The Taylor deformation parameter of the drop is determined by $D\equiv (L_d-l_d)/(L_d+l_d)$ where $L_d$ and $l_d$ are the long and short axis of the deformed drop respectively. An undeformed, spherical droplet corresponds to a deformation parameter $D=0$ while $D\rightarrow 1$ if the droplet is unboundedly stretched.

In its simplest form the different stages of deformation of a viscous drop in simple linear shear flow can be characterized by the Reynolds number $Re$ and capillary number $Ca$, respectively. For low Reynolds numbers $Re\rightarrow 0$, i.e.~in the Stokes flow limit, and for a viscosity ratio of $\lambda^{\ast}=1$, a critical capillary number $Ca^{*}\approx 0.42$ is found. For this particular viscosity ratio, the deformation of the droplet is approximately ellipsoidal until it reaches the limit $Ca^{*}$ where it breaks up into smaller droplets. For large viscosity ratios $\lambda^{\ast}\gtrsim 4$, the droplet reaches an maximum elongation $D^{max}$ and does not disintegrate as $Ca$ is further increased unless the Reynolds number $Re$ becomes comparable to unity~\cite{Rallison1984}.

Increase of the shear rate $\dot\gamma$ leads to an increase of the Reynolds number $Re \propto Ca$ where the ratio between $Re$ and $Ca$ is given by the Ohnesorge number $Oh^{-2}=Re/Ca$ which relates viscous to inertial forces. The critical capillary number $Ca^{*}$ of break up itself depends on $Re$ since the acceleration of the fluid particles in the rotating droplet and continuous fluid counteracts the stabilizing capillary stresses. Consequently, the critical capillary number is lowered $Ca^{*}(Re)<Ca^{*}(Re=0)$ limiting the range of stable, stationary droplet shapes as it has been numerically explored in Ref~\cite{Li2000}.

We simulated two benchmark scenarios where in one the viscosity ratio between the drop and the bulk fluid is $\lambda^{\ast}\approx 1$ and in the other $\lambda^{\ast}\approx4 $ with the following system parameters. For both benchmarks the initial radius of the drop is $R=8a$ and the drop is placed as a sphere in the center of the domain. The system size is $L=16R\times 8R \times 8R$ which is large enough to avoid any boundary effects~\cite{Janssen2007}. We apply again periodic boundary conditions in $x$-- and $y$--direction and Lees--Edwards boundary conditions in $z$--direction (see Sec.~\ref{ssec:viscosity}). The system temperature $T=5\times 10^{-3}$ and average particle number $\Navg=15$ is kept constant in all simulations. For $\lambda^{\ast}\approx 1$ all particles have mass $m=1$ whereas for $\lambda^{\ast}\approx 4$ the mass of the drop particles is increased to $m=5$ leading to an increase of viscosity by a factor of $\approx 4$ (see also Sec.~\ref{ssec:viscosity}). All systems are equilibrated for $2\times10^4$ time steps until the velocity field is stable and afterwards the drop shape is averaged over $5\times10^4$ subsequent time steps.

Shapes of the tank--treading drops in steady--state are summarized in Fig.~\ref{fig:taylor}. For small shear rates in panels (a) and (b) of Fig.~\ref{fig:taylor}, the shapes are almost perfectly ellipsoidal. Drops at larger shear rates with stronger deformations in panels (c) and (d) of Fig.~\ref{fig:taylor} tend to be more elongated. Deformation $D$ is plotted against the capillary number $Ca$ in Fig.~\ref{fig:taylor}e for  $\lambda^{\ast}\approx 1$ (blue symbols) and $\lambda^{\ast}\approx 4$ (red symbols). Additionally, the dashed line shows the linear relation for small droplet deformations from Refs.~\cite{Taylor1932,Taylor1934} given by
\begin{equation}
D = \frac{19\lambda^{\ast} +16}{16\lambda^{\ast}+16} Ca~,
\label{eq:taylor}
\end{equation}
for $\lambda^{\ast}=1$ \newtext{and $Re=0$}. The black symbols in Fig.~\ref{fig:taylor}e are taken from Ref.~\cite{Li2000} where the authors compare different numerical implementations of this problem for $\lambda^{\ast}=1$ and $Re=0$. For small capillary numbers up to $Ca\approx 0.1$ the $\MCSRDp$ simulations show the same linear relation as presented by other authors \cite{Taylor1934,Li2000}. If the capillary number is increased the $\MCSRDp$ simulations with $\lambda^{\ast}\approx 1$ and $\lambda^{\ast}\approx 4$ show a strong deformation already at relatively small capillary numbers. Furthermore, also the drop breakup occurs at smaller capillary numbers of $Ca\approx 0.24$. In the inset in Fig.~\ref{fig:taylor}e we show the Reynolds number $Re$ and capillary number $Ca$ for our two benchmark sets. The dashed line in the inset is taken from Ref.~\cite{Li2000} and marks a stability regime for the case $\lambda^{\ast}=1$ (blue symbols in our case) where drops are stable for small $Ca$ and small $Re$ (region left of the line). For small $Ca$ much larger values for $Re$ are needed before the drop breaks up into daughter droplets. Because in our $\MCSRDp$ simulations $Re \propto Ca$, the systems are not in the Stokes flow limit \newtext{of $Re=0$}. In the inset in Fig.~\ref{fig:taylor}e one can see that all but the last two of the data points for the case $\lambda^{\ast}\approx 1$ (blue circles) are below the breakup line. Considering the larger error bars especially for that last point (larger fluctuations) it may be that this drop also breaks up for longer simulation times. Blue crosses indicate drops that are unstable and have disintegrated into smaller droplets. A larger viscosity ratio (red symbols) counteracts increasing $Re$ and stable droplet shapes with stronger deformations are possible which is also indicated by a decreasing Ohnesorge number. This corresponds to the findings of e.g. Ref.~\cite{Taylor1934} that for $Re\ll1$ and $\lambda^{\ast} \gtrsim 4$ drop breakup no longer occurs no matter how large $Ca$ is.

This short example shows the capability of the $\MCSRDp$ algorithm to be used to study deforming droplets in linear shear flow \newtext{at finite $Re$ numbers}. Especially, the possibility to alter the viscosity ratio between the drop and the bulk fluid may be of interest for future studies.

\section{Surface wettability}
\label{sec:wetting}

Low capillary number flows of two or more immiscible phases in confinements are governed by the relative affinity of the phases to the walls. Numerical models for multi--phase flows in contact to the walls of, e.g. a microfluidic device or a porous medium have to capture not only complete wetting or non--wetting conditions, but also partial wetting. To this end we developed a method that accounts for relative adhesion of fluids to a solid surface and respects the no--slip boundary condition.

\subsection{Implementation}

Different affinities of the immiscible fluid phases to solid walls are controlled in our multi--color model through a virtual fluid phase inside the walls. These virtual fluid particles have been introduced already in Sec.~\ref{ssec:mpcmodel} to enforce the no--slip boundary condition for a mono--phase fluid in cases where the mean free path is small compared to the size of the collision cells. The adhesion strength of a droplet in contact to the wall can be controlled if we assign the virtual particles a certain color. Still, the virtual wall particles do not participate in the streaming step. Instead, they are removed and created anew after or before every collision step, respectively. Full wetting conditions are reached if we assign all virtual particles the color of the droplet phase. Likewise, the droplet is fully non--wetting if all virtual particles are assigned the color of the ambient fluid phase. Partial wetting of the fluid phases is achieved in a certain range of mixing ratios of virtual particles with colors of either the droplet or the continuous phase. Once a mechanical equilibrium is reached, the fluid--fluid interface of the droplet intersects the solid wall at a certain contact angle $\theta$. Besides a dependence on the control parameter $T$ and $\Navg$, the equilibrium contact angle will be a function of the color ratio
\begin{equation}
\lambda_{\mathrm{VP}} = \frac{N_{A}}{N_{A}+N_{B}}~,
\label{eq:ratio}
\end{equation}
where $N_{A}$ is the amount of virtual particles with the color of the bulk fluid and $N_{B}$ the amount of virtual particles with the color of the droplet, respectively. The corresponding contact angles $\theta$ need to be determined from a series of simulations.

\subsection{Static droplet on a homogeneously wetting substrate}
\label{ssec:static}

\begin{figure}
	\centering
	\includegraphics[width=0.8\columnwidth]{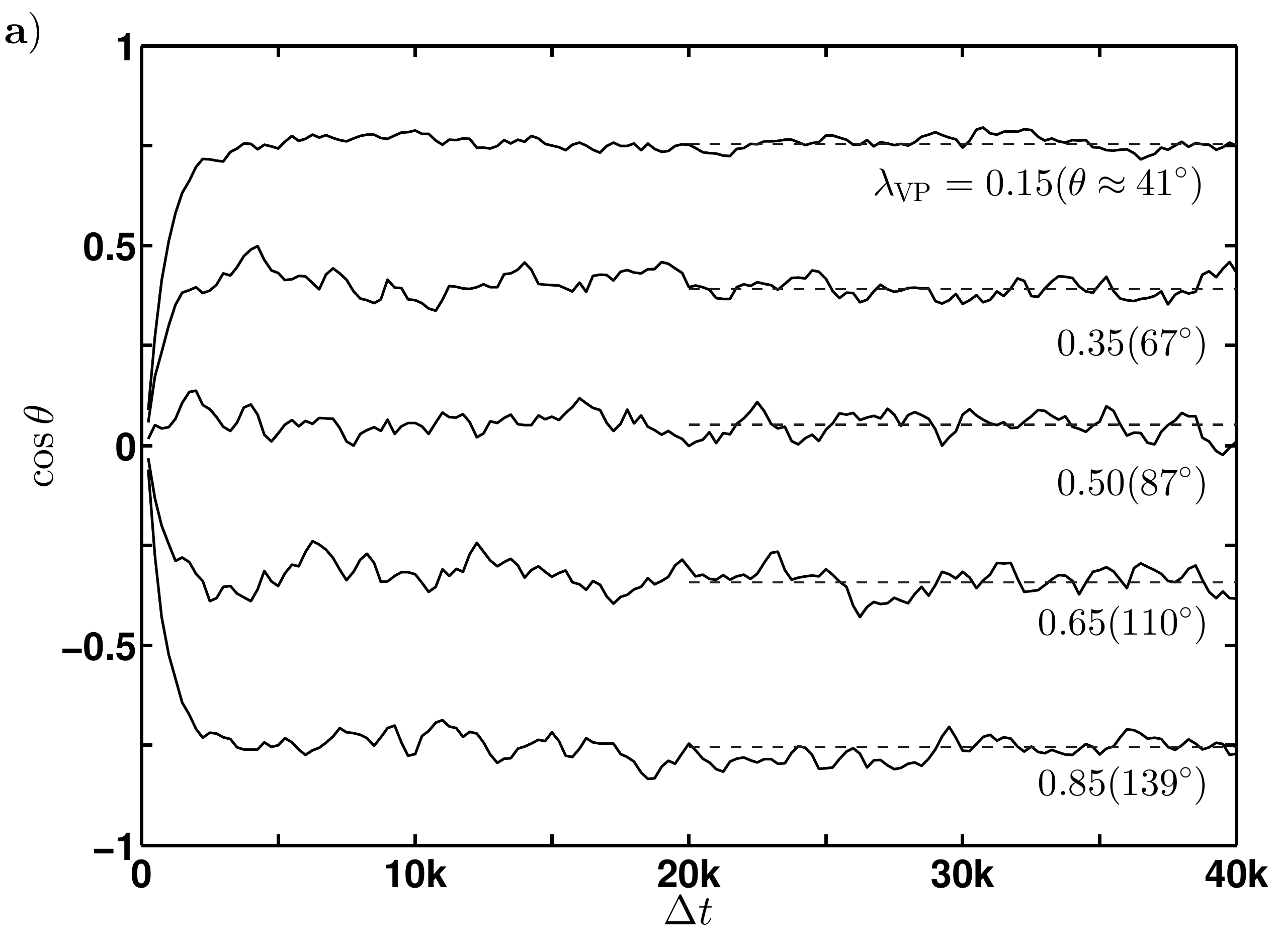}
	\includegraphics[width=0.8\columnwidth]{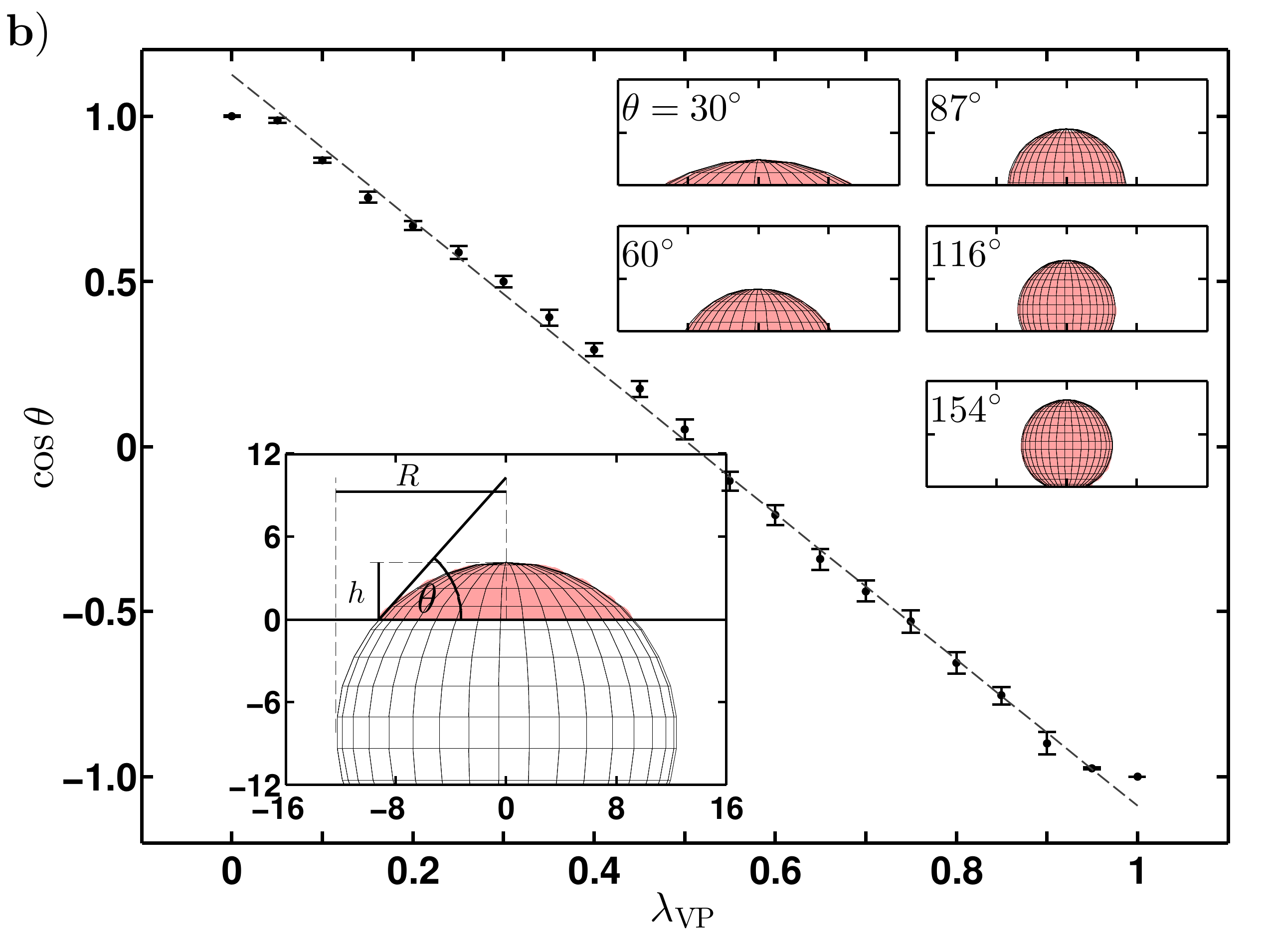}
	\caption{(color online) (\textbf{a}) evolution of the contact angle $\theta$ over time for five different ratios $\lambda_{\mathrm{VP}}$; the corresponding time averaged contact angles $\theta$ are indicated by the dashed lines; (\textbf{b}) cosine of the contact angle $\theta$ as a function of the ratio $\lambda_{\mathrm{VP}} = N_{A}/(N_{A}+N_{B})$ (dashed line as a guide to the eye); the large inset illustrates the determination of $\theta$ by fitting a spherical cap to the time averaged droplet interface; the small insets show time averaged droplet interfaces (red) and corresponding spherical cap fits (grid overlay) for five different contact angles.}
	\label{fig:contact}
\end{figure}

To determine the dependence of the contact angle $\theta$ on the ratio $\lambda_{\mathrm{VP}}$ we perform the following experiment. In a cubic simulation box of size $L^3=32^3$ with periodic boundary conditions in $x$-- and $y$--direction and bounce--back boundary conditions in $z$--direction a spherical cap--shaped droplet is placed at the center of the lower $z$--surface at $L_{z}=0$ with a radius of $R=7a$. The temperature of the system is $T=5\times 10^{-3}$ and the average particle number is $\Navg=20$. Because the average particle number in the simulation box is the same as inside the walls it is possible to measure 21 configurations of a partially wetting droplet in mechanical equilibrium. The contact angle $\theta$ is measured in the following manner. The time averaged droplet interface is fitted with a spherical cap to determine the radius $R$ and the height $h$ (see inset in Fig.~\ref{fig:contact}b). From these fitted values $\theta$ is calculated with
\begin{equation}
	\cos{\theta} = \frac{R-h}{R}
\label{eq:theta}
\end{equation}
Figure~\ref{fig:contact}a exemplifies the evolution of the contact angle $\theta$ over time for five different ratios of $\lambda_{\mathrm{VP}}$. The equilibrium contact angle is already reached after a maximum of $2\:000$ time steps. After this initial time span the contact angle is stable within small fluctuations for the remainder of the simulation. Figure~\ref{fig:contact}b shows the cosine of the final contact angle as a function of $\lambda_{\mathrm{VP}}$. The plotted value is a time average over $2\times 10^{4}$ time steps taken after $2\times 10^{4}$ time steps of equilibration of the data shown in Fig.~\ref{fig:contact}a (dashed lines to the individual curves). The error bars show the standard deviation of the contact angle values over the averaging period. Other than for the two extreme cases where $\lambda_{\mathrm{VP}}=0$ and $\lambda_{\mathrm{VP}}=1$ the relationship between the contact angle and $\lambda_{\mathrm{VP}}$ is broadly linear. The small insets in Fig.~\ref{fig:contact}b show the time averaged droplet interface (red) and the corresponding fit of the spherical cap (grid) for five different realizations. Note the remarkably well overlap of the interface and the fitted spherical cap. For the extreme cases $\lambda_{\mathrm{VP}}=0$ and $\lambda_{\mathrm{VP}}=1$, as expected, the droplet either completely wets the surface ($\cos{\theta_{0}}=1$) or detaches from the surface ($\cos{\theta_{1}}=-1$).

\subsection{Droplet dewetting from a stripe geometry}
\label{ssec:stripe}

\begin{figure}
	\centering
	\includegraphics[width=0.49\columnwidth]{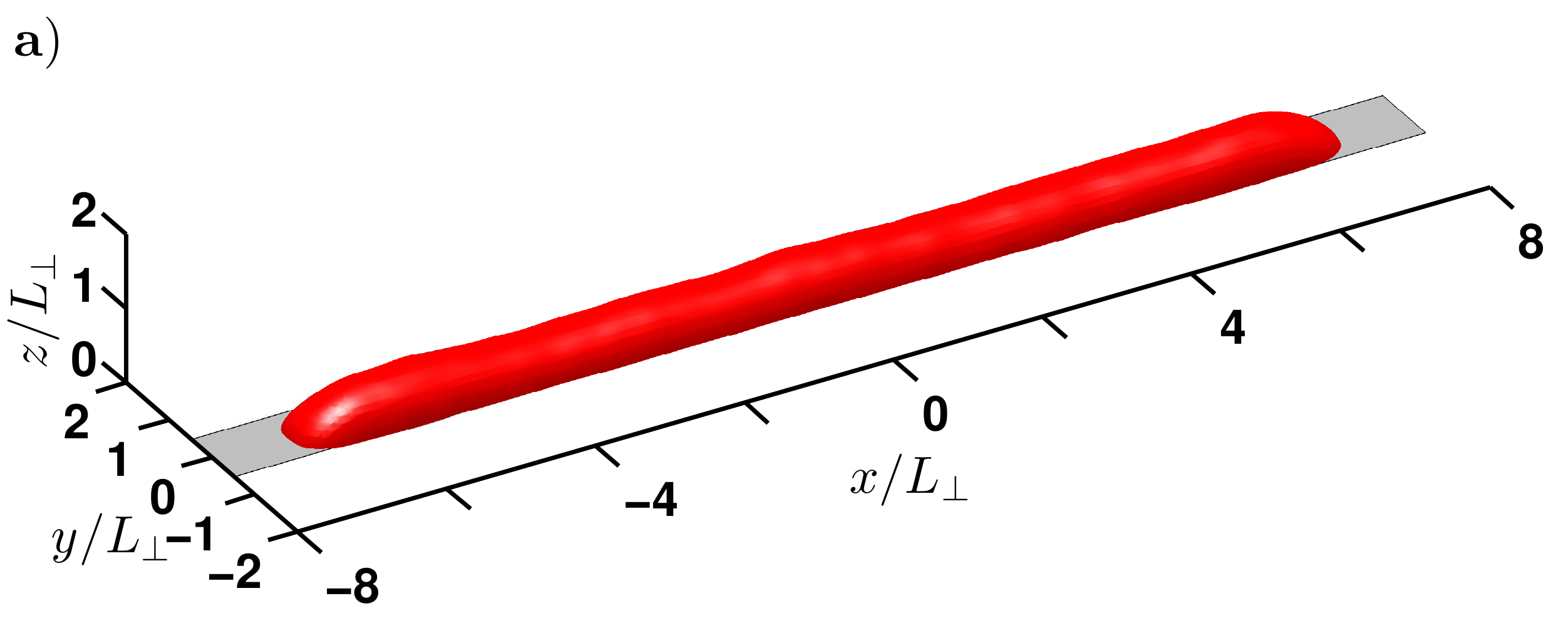}
	\includegraphics[width=0.49\columnwidth]{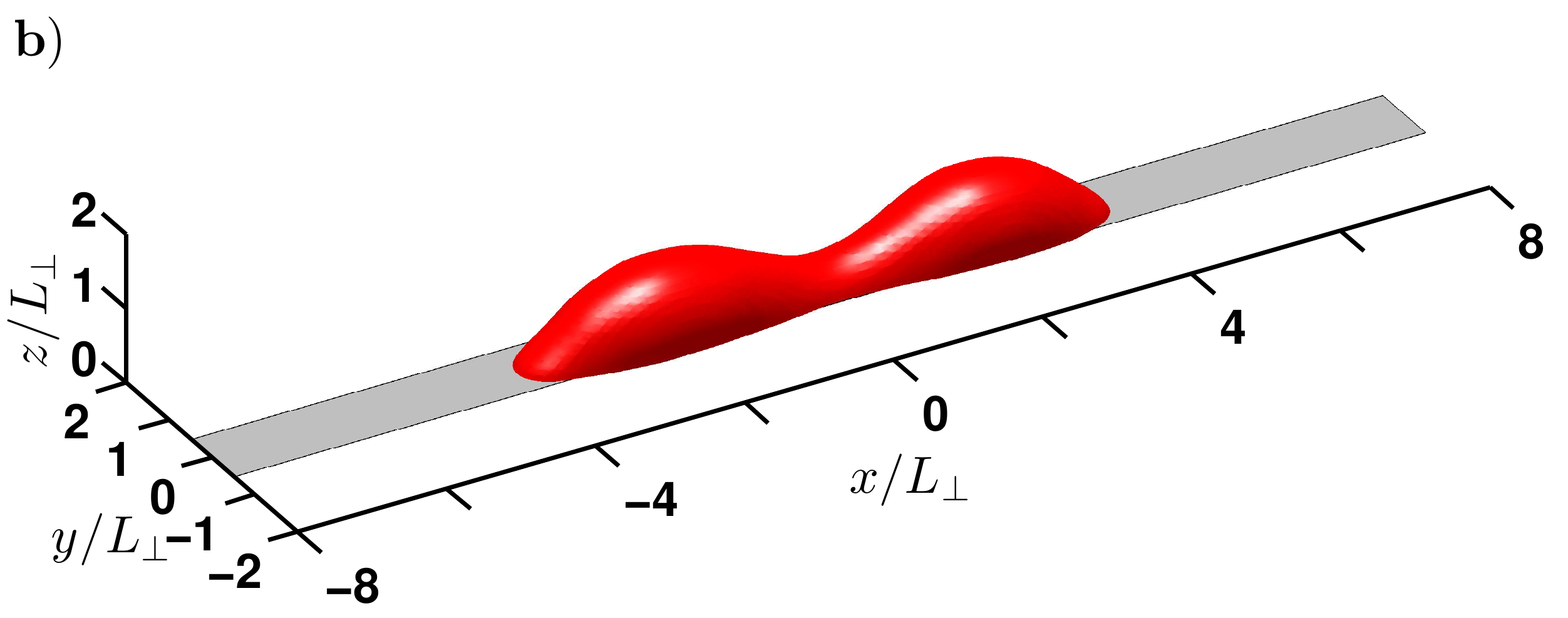}
	\includegraphics[width=0.49\columnwidth]{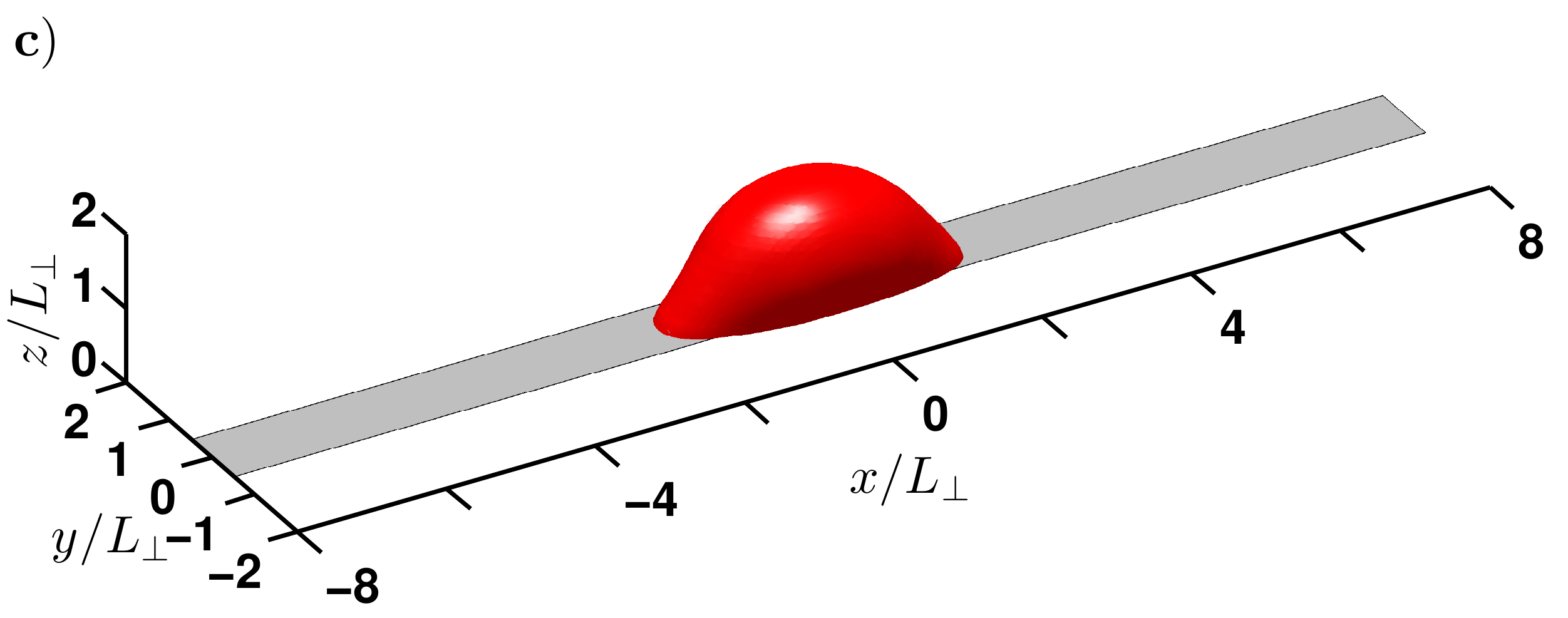}
	\includegraphics[width=0.49\columnwidth]{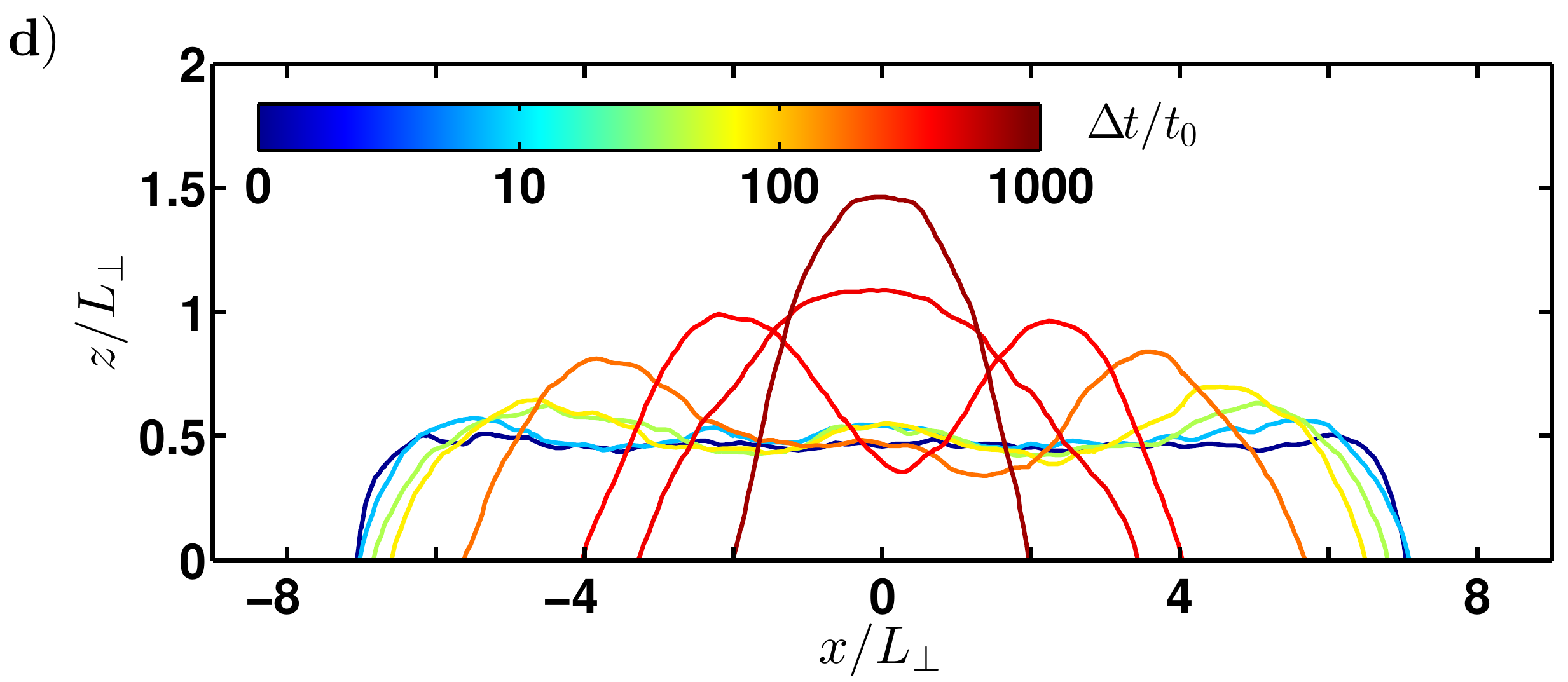}	
	\caption{(color online) all spatial units are rescaled by the stripe width $L_\perp$ and time is rescaled by the viscous capillary time scale $t_{0}=\eta L_{\perp}/\gamma\approx126.3$ (\textbf{a}--\textbf{c}); snapshots of the temporal evolution of a droplet interface shape on a wetting stripe; the volume of the droplet is $V\approx5L_{\perp}^3$ and the contact angle on the stripe is $\theta_{0}=54^\circ$; (\textbf{d}) interface profiles for the same droplet in the center of the stripe, vertical exaggeration by a factor of 9.}
	\label{fig:stripe}
\end{figure}

As a further benchmark to test whether the interfacial flows in contact to solid walls are faithfully reproduced in our $\MCSRD$ implementation, we study the dewetting of a liquid droplet from a wettability pattern \cite{Gau1999,Lipowsky2000,Brinkmann2002,Klingner2004}. A linear stripe of high wettability is created using spatial modulation of the color ratio of virtual wall particles in a rectangular region of a plane solid wall. A series of color ratios of virtual particles is chosen to achieve a low contact angle $\theta_0<90^\circ$ on the stripe,  while the high contact angle on the surrounding surface is fixed to $\theta_m=180^\circ$. In the present study we consider a ratio $L_\parallel/L_\perp=42$ of stripe length $L_\parallel=256a$ to stripe width $L_\perp=8a$. A large simulation box  with dimensions $256\times 32 \times 32$ and periodic boundary conditions in $x$-- and $y$--direction ensures that the shape evolution of the dewetting droplets are not affected by the finite size of the system. Initially, a flat cylindrical droplet of varying volume in the range of $V=1L^{3}_{\perp}$ to $V=10L^{3}_\perp$ is deposited on the stripe. Here, we chose the length of the droplet to be shorter than the stripe to avoid a connection of the wetting fluid phase across the periodic boundary. As a result, a fraction at the end of the stripe remains in contact to the ambient non--wetting fluid. 

Figure~\ref{fig:stripe} illustrates the evolution of a single droplet during different stages of the dewetting process for a contact angle of $\theta_{0}=54^\circ$ on the stripe and $\theta_{m}=180^\circ$ on the surrounding substrate. The initial length and height of the wetting droplet are chosen to be $l_0=13.75L_{\perp}$ and $h_0=0.375L_{\perp}$ with a droplet volume of $V\approx5L_{\perp}^3$. Following the evolution of shapes, one can clearly see that the dewetting process starts at the end of the filamentous droplet and progresses towards the droplet's center. Because of mass conservation, the wetting fluid on the stripe is \emph{piled up} behind the inward moving contact line. As expected from the wettability contrast between the stripe and the surrounding matrix, the lateral parts of the contact line remain pinned to the side of the stripe. Only at the end of the dewetting process, and for sufficiently large volume of the droplets, the lateral part of the contact line depinns from the stripe edges and displays excursions onto the surrounding non--wetting matrix.

\begin{figure}
	\centering
	\includegraphics[width=0.85\columnwidth]{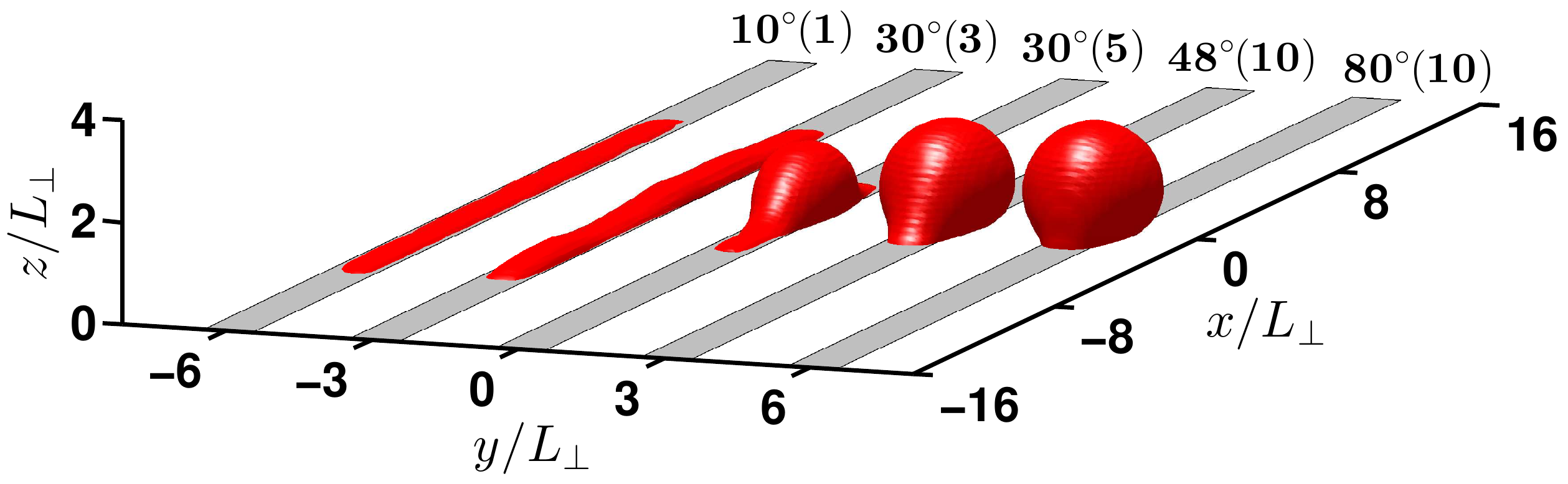}	
	\caption{(color online) all spatial units are rescaled by the stripe width $L_\perp$; different final droplet configurations on a wetting stripe with varying droplet volumes $V$ \newtext{(the value in brackets)}, the individual droplets are joined in one plot for visibility reasons.}
	\label{fig:stripe2}
\end{figure}

Final droplet shapes for a series of different contact angles on the stripe and different droplet volumes are shown in Fig.~\ref{fig:stripe2}. Besides a small \emph{snail foot} on the stripe, the final shape of the liquid interface is close to a spherical cap. Only for small contact angles on the stripe smaller than $\theta=30^\circ$ and corresponding small volumes $V$, the dewetting process stops before the spherical droplet shapes are reached (two leftmost configurations). In these cases, the droplet relaxes into a spread--out, filamentous shape. In most parts droplets of the latter class display a homogeneous cross section. Deviations from the cylindrical shape are localized to the rounded end caps scaling with the width of the stripe. 

Both interfacial morphologies, the spherical and the cylindrical droplet shapes compare well to the numerical energy minimizations reported in Ref.~\cite{Brinkmann2002}. In agreement with the predictions of Ref.~\cite{Brinkmann2002}, we find spread--out, filamentous morphology for small contact angles $\theta\lesssim 39^\circ$ and droplet--like compact shapes for contact angles $\theta\gtrsim 39^\circ$ when considering the corresponding droplet volume $V$.

\section{Conclusion}
\label{sec:conclusions}

In this work we extended the multi--color SRD algorithm for immiscible fluid phase flow of Inoue et al.~\cite{Inoue2004} to include general wetting conditions of the walls. An additional modification of the SRD collision operator guarantees local conservation of vorticity and thus avoids artifacts in interfacial flows for a viscosity ratio of the fluids that differs from unity. To demonstrate the versatility of our simulation method, we conducted systematic measurements of the dynamic shear viscosity and interfacial tension, and performed a number of benchmarks for interfacial flows.
 
Within the relevant range of simulation parameters average particle number $\Navg$ and temperature $T$, the multi--color collision operator acts similar to a standard SRD collision operator with a fixed collision angle $\alpha$. The dynamic shear viscosity of a mono--phase fluid in the multi--color SRD model agrees well with the viscosity of the standard SRD model for certain collision angles $\alpha$. In particular, we find equivalent collision angles $\alpha_{\rm E} \simeq 90^\circ$ and $\alpha_{\rm E}\simeq 60^\circ$ for the SRD model without and with angular momentum conservation, respectively. Explicit measurements of the local stress in the SRD fluid subject to a linear shear flow show the expected symmetry of the stress tensor components whenever local vorticity conservation is respected in the collisions.

Local stresses in the fluids are measured by an \oldtext{areal}\newtext{area--weighted} averaging of the linear momentum flux in a fine grid of control surfaces. This method turned out to be particularly useful in measurements of the stress profile across the interface between two immiscible fluid phases. A comparison to corresponding stress profiles from volume averages derived from the virial theorem reveals an inconsistency of the latter method for SRD simulation methods which can be understood from the non--local exchange of linear momentum in the collision cells and the ambiguity of stress localization. 

The interfacial tension derived from the stress profile is further validated with corresponding values obtained from two independent methods. Exploiting the equation of state of the SRD fluid, being that of an ideal gas, we can simply relate the pressure difference between the drop and the ambient fluid phase to the difference of particle densities, and obtain the value of the interfacial tension from the Young--Laplace relation. In thermal equilibrium, we can employ the equipartition theorem to relate the amplitudes of thermally excited capillary waves to the magnitude of interfacial tension. Measurements of the power spectr\oldtext{um}\newtext{a} confirm\oldtext{s} the expected power law decay and prefactor. The values of the interfacial tension obtained from all three methods turn out to be in very good agreement. 

Deformations of a viscous drop that is subject to a linear shear flow are governed by both the dynamic shear viscosity and the interfacial tension, and was chosen therefor as a benchmark to validate the correct interplay of capillary and viscous stresses. In the limit of small capillary numbers, the angular momentum conserving multi--color operator correctly reproduces the deformation of the immersed droplet, as predicted by Taylor~\cite{Taylor1932,Taylor1934} for the viscous--capillary limit of small shear rates. Regions of stable and unstable drop configurations are comparable to results published by other authors within the range of Capillary and Reynolds numbers studied.

Assigning colors not only to the fluid particles but also to the virtual particles in the walls gives us the possibility to model different affinities of the fluids to the wall. Virtual wall particles were initially proposed to achieve a no--slip boundary condition for densities and temperatures where the mean free path of the fluid particles is small compared to the collision cells. For simplicity, we considered mixtures of wall particles with colors corresponding to the two bulk phases. Varying the color ratio of the wall particles allows us to control the adhesion of fluids to the walls. The corresponding contact angles between a complete wetting and a non--wetting situation were obtained from fits to the shapes of equilibrated sessile drops. Our extended multi--color SRD model also reproduces the effects of a spatially varying wall wettability onto the equilibrium shapes of sessile drops and interfacial flows. As a benchmark, we considered the well studied case of a liquid drop adhering to a plane wall decorated with a wettable stripe on an otherwise non--wettable surface. Shapes of equilibrated drops are consistent with corresponding shapes recorded in wetting experiments and model calculations and the dynamics of the free interface during the transitions conform to expectations for interfacial flows with small slip length.

In summary, we have shown that our extended multi--color SRD algorithm provides a useful tool to study a wide range of fluid mechanics problems that involve adhesion of immiscible fluid phases to solid walls. These could be for example the imbibition of a fluid into a porous media filled with another fluid that exhibits a different wettability to the porous matrix, a situations that is encountered in many porous rocks and thus relevant for reservoir engineering.
Additionally, the walls of the porous medium itself could exhibit certain patterns of differently wettable walls. Our model should also be of use for a number of applications in micro-- or nanofluidics. The possibility to define an arbitrary number of mutually immiscible drop phases which all interact identically with the ambient fluid and the walls opens the possibility to study flows of emulsion droplets. The interplay between the involved fluids and the confining walls is important and can now be studied by means of stochastic rotation dynamics.

\section*{Acknowledgments}
The authors acknowledge helpful discussions with Stephan Herminghaus, Marco G. Mazza, Badr Kaoui and Julie Murison. Generous support was granted from the Exploratory Research (ExploRe) program of BP Plc.

\section*{References}
\bibliographystyle{elsarticle-num}

\end{document}